\documentclass[10pt,a4paper]{article}
\pdfoutput=1

\usepackage{amsmath}
\usepackage{amsfonts}
\usepackage{amssymb}
\usepackage[pdftex]{color,graphicx}

\begin{document}

\title{\textbf{Charged black holes in string-inspired gravity}\\
\textsf{\large{I. Causal structures and responses of the Brans-Dicke field}}}

\author{\textsc{Jakob Hansen$^{a}$}\footnote{{\tt hansen{}@{}kisti.re.kr}}\;\;
and \textsc{Dong-han Yeom$^{b,c}$}\footnote{{\tt innocent.yeom{}@{}gmail.com}}\\
\textit{$^{a}$\small{KISTI, Daejeon 305-806, Republic of Korea}}\\
\textit{$^{b}$\small{Yukawa Institute for Theoretical Physics, Kyoto University, Kyoto 606-8502, Japan}}\\
\textit{$^{c}$\small{Leung Center for Cosmology and Particle Astrophysics,}}\\
\textit{\small{National Taiwan University, Taipei 10617, Taiwan}}
}

\maketitle

\begin{abstract}
We investigate gravitational collapses of charged black holes in string-inspired gravity models, including dilaton gravity and braneworld model, as well as $f(R)$ gravity and the ghost limit. If we turn on gauge coupling, the causal structures and the responses of the Brans-Dicke field depend on the coupling between the charged matter and the Brans-Dicke field. For Type~IIA inspired models, a Cauchy horizon exists, while there is no Cauchy horizon for Type~I or Heterotic inspired models. For Type~IIA inspired models, the no-hair theorem is satisfied asymptotically, while it is biased to the weak coupling limit for Type~I or Heterotic inspired models. Apart from string theory, we find that in the ghost limit, a gravitational collapse can induce inflation by itself and create one-way traversable wormholes without the need of other special initial conditions.
\end{abstract}

\begin{flushright}
{\tt YITP-14-47}
\end{flushright}

\newpage

\tableofcontents

\newpage

\section{Introduction}

To investigate the theory of everything which explains gravity and quantum theory in a consistent way, several candidates have been proposed \cite{DeWitt:1967yk,Kiefer} with string theory being one of the most competitive ideas \cite{Green:1987sp}. String theory can explain not only the finite quantum interactions of gravitons but also the possible modifications of gravity and matter sectors \cite{Becker:2007zj}. Such modified gravity models or additional matter contents are useful to study various topics, for example inflationary and late time cosmology \cite{Kachru:2003aw}, black hole physics and applications to AdS/CFT \cite{Maldacena:1997re} or astrophysical applications \cite{Conlon:2010jq} etc. Moreover, if string theory is the correct model of quantum gravity then such modifications should explain various traditional problems of gravity, e.g., the singularity problem of black holes \cite{Hawking:1969sw} or cosmology \cite{Borde:2001nh}.

A typical modification via string theory is the existence of a dilaton field \cite{Gasperini:2007zz}. This is in fact the same as the Brans-Dicke field \cite{Brans:1961sx} which controls the interaction coupling strengths. The kinetic term and the potential term of the Brans-Dicke field can be changed by a conformal transformation from the string frame (or the Jordan frame) to the Einstein frame and is equivalent to a canonical scalar field model in the Einstein frame \cite{Fujii:2003pa}. However, if the model couples to other interactions, then the matter sector is no longer canonical after the conformal transformation, an interesting example of which is the gauge field. In Section~\ref{sec:gen}, we summarize various types of string effective actions and the different types of interactions they generate between the dilaton field and the gauge field.

In this paper, we want to investigate responses of the Brans-Dicke field that depend on couplings between the Brans-Dicke field and the gauge field. Naively speaking, they will form charged black holes and there are already a lot of papers which investigate stationary solutions \cite{Gibbons:1987ps}. However, in general the dynamics will be complicated for realistic gravitational collapses. For such cases we expect complicated phenomena such as :
\begin{itemize}
\item[--] \textit{Causal structure :} In the general string frame (Jordan frame), the null energy condition can be violated and hence the dynamics of locally defined horizons \cite{Ashtekar:2004cn} can be complicated. In addition, the existence of the central singularity and the Cauchy horizon curvature singularity \cite{Poisson:1990eh} should be affected by the dynamics of the Brans-Dicke field.
\item[--] \textit{Response of the Brans-Dicke field :} Already some solutions are known which violate the assumptions of the no-hair theorem \cite{Israel:1967wq,Gibbons:1987ps}. However, is it be possible to form scalar hairs from gravitational collapses? For neutral black holes it was investigated by \cite{Yeom2}, but the responses due to the charge can cause quite different phenomena. This may be related to the destabilization of the moduli/dilaton field that should be controlled by a proper potential and may have observational implications \cite{Conlon:2010jq}.
\end{itemize}
We will mainly focus these two topics by using numerical techniques applied to the double-null formalism. To the best knowledge of the authors, the double-null formalism is one of the most efficient techniques to investigate the interior of the black holes \cite{Hamade:1995ce} and we have priviously used this approach with good success \cite{authors, Hansen:2013vha, JHY}. There are some previous works regarding string-inspired models using the double-null formalism \cite{doublenull,Ann} and other methods \cite{Scheel:1994yr} as well as the authors' previous works \cite{authors,Yeom1,Yeom3}. In this paper we will investigate a wide parameter space, for example, we will cover not only directly string-inspired models, but also indirectly string-inspired models ($f(R)$ inspired model \cite{Yeom5}) as well as natural generalizations of the parameters, e.g., the ghost limit.

In fact, our potential parameter space is huge and can give rise to a wide range of setups, for example :
\begin{itemize}
\item[--] \textit{Model dependence}: Type~I, Type~II, Heterotic, or braneworld inspired models, etc.
\item[--] \textit{Potential dependence}: whether there is a potential that stabilizes the moduli or dilaton field or not \cite{Conlon:2010jq}.
\item[--] \textit{Asymptotic conditions}: de Sitter, Minkowski, or anti de Sitter \cite{Hansen:2013vha,Hwang:2010aj}.
\item[--] \textit{Symmetries}: spherical, planar, hyperbolic, or other symmetries \cite{Hansen:2013vha,Yeom4}.
\item[--] \textit{Dimensions}: 3, 4, or $D$ dimensions \cite{Hwang:2010aj}.
\end{itemize}
Although some of these topics were already investigated by the authors and others, still there are many interesting and various topics to be studied. In this paper, we will limit our focus to the basic model dependence with no dilaton potential, asymptotic flat, spherical symmetry and four dimensions, hence the paper's subtitle : ``\textit{I. Causal structures and responses of the Brans-Dicke field.}''. To investigate more general features of mass inflation (for a basic review of mass inflation, see Appendix~A and references therein), we need to investigate the cases with different potentials as well as various parameters. Furthermore, it is interesting to change the asymptotic cosmological constant and different symmetries. The variation of dimensions and their semi-classical responses will be also interesting to study. These issues are beyond the scope of this paper, but will be investigated in future works.

Apart from the physics involved, this paper is by itself important by it's technical merit, since this is a quite generalized and complicated implementation of the double-null formalism. We are sure that there are many more variations to be made from this numerical technique and that this paper is an important step towards further useful or realistic string-inspired models. We hope that this paper can be a leading paper that will introduce upcoming papers.

This paper is organized as follows; In Section~\ref{sec:gen}, we summarize various models inspired from string theory. In Section~\ref{sec:mod}, we present our model in the double-null formalism. In Section~\ref{sec:dyn}, we present and summarize two of our main results, specifically the issue of causal structures (e.g., dynamics of local horizons and existence of Cauchy horizons) and responses of the Brans-Dicke field (e.g., check the no-hair theorem and the direction of coupling bias). In Section~\ref{sec:dis}, we summarize and discuss future issues. Finally, in the Appendices, we comment on basics of mass inflation (Appendix A), the consistency and convergence of our numerical scheme (Appendix B) and present a catalog of simulations for various initial conditions (Appendix C).

\section{\label{sec:gen}General review of motivations}

In this paper, we consider the following action ($c=G=\hbar = 1$):
\begin{eqnarray}\label{eq:BDscalar_or}
S = \frac{1}{16\pi} \int \sqrt{-g}d^{4}x \left[ \Phi R - \frac{\omega}{\Phi}\left( \nabla \Phi \right)^{2}  -\Phi^{\beta}F_{\mu\nu}^{2} \right],
\end{eqnarray}
where $\Phi$ is the Brans-Dicke field, $R$ is the Ricci scalar, $\omega$ is the Brans-Dicke coupling, $F_{\mu\nu}$ is the kinetic term of the gauge field and $\beta$ is the coupling between the Brans-Dicke field and the gauge field. As we will see below, $\omega$ and $\beta$ in particular are model depent.

Of course, this model is limited up to order $R$ and hence does not correspond to real string theory, since the real theory should contain all possible interactions, including higher order curvature corrections. On the other hand, this model is sufficiently general to cover a lot of string inspired models and thus provide important intuitions concerning dynamical black hole physics.

Next, we illustrate some of the models and motivations which is covered by our action (Equation~(\ref{eq:BDscalar_or})).

\paragraph{Dilaton gravity}

Every low energy effective action of string theory contains the following sector that includes the dilaton field \cite{Gasperini:2007zz}:
\begin{eqnarray}
\label{eq:dilaton} S = \frac{1}{2 \lambda_{s}^{d-1}}\int d^{d+1}x \sqrt{-g} e^{-\phi} \left( R + (\nabla \phi)^{2} \right),
\end{eqnarray}
where $d$ is the space dimensions, $\lambda_{s}$ is the length scale of string units, and $\phi$ is the dilaton field. This can be transformed by redefining $\phi$, such as
\begin{eqnarray}
\label{eq:def} \frac{e^{-\phi}}{\lambda_{s}^{d-1}} = \frac{\Phi}{8 \pi G_{d+1}},
\end{eqnarray}
where $G_{d+1}$ is the $d+1$ dimensional gravitation constant. Then, we recover the Brans-Dicke theory in the $\omega = -1$ limit.

In general, there will be higher order loop corrections to the low energy effective action. For example, in a model from Heterotic string theory compactified on a $Z_{N}$ orbifold \cite{Foffa:1999dv}, this can be well approximated by the correction to the Brans-Dicke coupling $\omega$:
\begin{eqnarray}
S = \frac{1}{2 \lambda_{s}^{2}}\int d^{4}x \sqrt{-g} e^{-\phi} \left[ R + \left(1+e^{\phi} \left( \frac{3\kappa}{2} \right) \frac{6+\kappa e^{\phi}}{(3+\kappa e^{\phi})^{2}} \right) (\nabla \phi)^{2} \right],
\end{eqnarray}
where $\kappa$ is a positive constant of order one which is related to anomaly coefficients. Therefore, the coupling should be field dependent and hence, quite large ranges of $\omega$ can be covered by string-inspired models.

\paragraph{Couplings to the gauge field:} When we further consider the coupling to the gauge field, the parameter $\beta$ may depend on the models \cite{Becker:2007zj}. We illustrate some detailed examples;
\begin{description}
\item[Type~IIA] The effective action (bosonic sector) of the Type~IIA is expanded by
\begin{eqnarray}
S_{\mathrm{IIA}} = \frac{1}{2 \lambda_{s}^{8}} \int d^{10}x \left[ \sqrt{-g} e^{-\phi} \left( R + (\nabla \phi)^{2} - \frac{1}{12} H_{3}^{2} \right) - \left( \frac{1}{4}F_{2}^{2} + \frac{1}{48}\tilde{F}_{4}^{2} \right) \right] + ...,
\end{eqnarray}
where $H_{3}$ is the field strength tensor of the NS-NS two-form field $B_{2}$, $F_{2}$ is the field strength tensor of the R-R one form field $A_{1}$ and $\tilde{F}_{4} = dA_{3} + A_{1} \wedge H_{3}$, where $A_{3}$ is the R-R three-form field. Therefore, after the dimensional reduction, it is reasonable to obtain the following effective action:
\begin{eqnarray}
S^{(4)}_{\mathrm{IIA}} = \frac{1}{16 \pi} \int d^{4}x \sqrt{-g} \left[ e^{-\phi} \left( R + (\nabla \phi)^{2} \right) - F_{2}^{2} \right] + ...,
\end{eqnarray}
and after the redefinition of dilaton, we obtain the gravitational sector by
\begin{eqnarray}
\propto \Phi \left( R + \frac{\left(\nabla \Phi\right)^{2}}{\Phi^{2}} \right)
\end{eqnarray}
and the two-form field term becomes
\begin{eqnarray}
\propto F_{2}^{2}.
\end{eqnarray}
Therefore, Type~IIA corresponds to the $\omega = -1$ and $\beta = 0$ case.

\item[Type~I] The effective action of Type~I is expanded by
\begin{eqnarray}
S_{\mathrm{I}} = \frac{1}{2 \lambda_{s}^{8}} \int d^{10}x \sqrt{-g} \left[ e^{-\phi} \left( R + (\nabla \phi)^{2} \right) - \frac{1}{12} \bar{H}_{3}^{2} - \frac{1}{4} e^{-\phi/2} \mathrm{Tr} F_{2}^{2} \right] + ...,
\end{eqnarray}
where $\bar{H}_{3}$ is the mixed contribution of the R-R two-form $A_{2}$ and of the matrix valued one-form $A_{1}$, $F_{2}$ is the field strength tensor of $A_{1}$ with the gauge symmetry by the $SO(32)$ group.
Therefore, after the dimensional reduction and the symmetry breaking to $U(1)$, we can reduce the action by
\begin{eqnarray}
S^{(4)}_{\mathrm{I}} = \frac{1}{16 \pi} \int d^{4}x \sqrt{-g} \left[ e^{-\phi} \left( R + (\nabla \phi)^{2} \right) - e^{-\phi/2} F_{2}^{2} \right] + ...;
\end{eqnarray}
and after the field redefinition, we obtain the same gravitational sector plus the two-form field term
\begin{eqnarray}
\propto \Phi^{1/2} F_{2}^{2}.
\end{eqnarray}
Therefore, Type~I corresponds to the $\omega = -1$ and $\beta = 0.5$ case.

\item[Heterotic] The effective action of the Heterotic theory is expanded by
\begin{eqnarray}
S_{\mathrm{het}} = \frac{1}{2 \lambda_{s}^{8}} \int d^{10}x \sqrt{-g} e^{-\phi} \left[ R + (\nabla \phi)^{2} - \frac{1}{2}\bar{H}_{3}^{2} - \frac{1}{4} \mathrm{Tr} F_{2}^{2} \right] + ...,
\end{eqnarray}
where $\bar{H}_{3}$ is the mixed contribution of the NS-NS two-form field $B_{2}$ and of the matrix valued one-form $A_{1}$, $F_{2}$ is the field strength tensor of $A_{1}$ with the gauge symmetry by the $SO(32)$ or $E_{8} \times E_{8}$ group.
Therefore, after the dimensional reduction and symmetry breaking to $U(1)$, we can reduce the action by
\begin{eqnarray}
S^{(4)}_{\mathrm{het}} = \frac{1}{16 \pi} \int d^{4}x \sqrt{-g} e^{-\phi} \left[ R + (\nabla \phi)^{2} - F_{2}^{2} \right] + ...;
\end{eqnarray}
and after the field redefinition, we obtain the same gravitational sector plus the two-form field term
\begin{eqnarray}
\propto \Phi F_{2}^{2}.
\end{eqnarray}
Therefore, Heterotic theory corresponds to the $\omega = -1$ and $\beta = 1$ case.
\end{description}

\paragraph{Braneworld models}

In the Randall-Sundrum model \cite{Randall:1999ee}, they introduced two branes, where two branes are connected by a warp factor. In the end, one side has a positive tension and the other side has a negative tension. According to Garriga and Tanaka \cite{Garriga:1999yh}, one can calculate the effective action on branes in the weak field limit as a Brans-Dicke-type theory and we obtain the $\omega$ parameter
\begin{eqnarray}
\label{eq:gt} \omega = \frac{3}{2} \left( e^{\pm s/l} - 1 \right),
\end{eqnarray}
where $s$ is the distance between branes, $l=\sqrt{-6/\Lambda}$ is the length scale of the anti de Sitter space and the sign $\pm$ denotes the sign of the tension. These braneworld inspired models can allow various range of $\omega$ near $-3/2$.

\paragraph{$f(R)$ gravity}

In general, string theory can introduce various higher curvature terms. If we mainly focus on the Ricci scalar sectors, then we obtain so-called $f(R)$ gravity. The action of $f(R)$ gravity is
\begin{eqnarray}
S = \frac{1}{16\pi} \int d^{4}x \sqrt{-g} f(R).
\end{eqnarray}
We can modify this action to the scalar-tensor type model by introducing an auxiliary field $\psi$. Then, the gravity sector changes the form \cite{Sotiriou:2008rp} as follows
\begin{eqnarray}
S = \frac{1}{16\pi} \int d^{4}x \sqrt{-g} \left[f(\psi) + f'(\psi) (R-\psi) \right],
\end{eqnarray}
where we have a constraint $\psi = R$. From this, by defining a new field $\Phi \equiv f'(\psi)$, we can rewrite the action
\begin{eqnarray}
S = \frac{1}{16\pi} \int d^{4}x \sqrt{-g} \left[\Phi R - V(\Phi) \right],
\end{eqnarray}
where
\begin{eqnarray}
V(\Phi) = - f(\psi) + \psi f'(\psi).
\end{eqnarray}
Now we obtain the Brans-Dicke theory in the $\omega = 0$ limit with a potential of the Brans-Dicke field, though we ignore the potential term in this paper.

\section{\label{sec:mod}Model for charged black holes}

\subsection{Brans-Dicke theory with $U(1)$ gauge field}

The prototype action of the Brans-Dicke theory with a $U(1)$ gauge field becomes
\begin{eqnarray}\label{eq:BDscalar}
S &=& \int \sqrt{-g}d^{4}x \left[\frac{1}{16\pi} \left( \Phi R - \frac{\omega}{\Phi}\Phi_{;\mu}\Phi_{;\nu}g^{\mu\nu} \right)\right. \nonumber \\
&&\left.+ \Phi^{\beta} \left(- \frac{1}{2}\left(\phi_{;\mu}+ieA_{\mu}\phi \right)g^{\mu\nu}\left(\overline{\phi}_{;\nu}-ieA_{\nu}\overline{\phi}\right)-\frac{1}{16\pi}F_{\mu\nu}F^{\mu\nu} \right) \right],
\end{eqnarray}
where $\phi$ is a complex scalar field\footnote{We abuse $\phi$ again; from now this is not the dilaton field but the complex scalar field.} with a gauge coupling $e$, $A_{\mu}$ is a gauge field, and $F_{\mu\nu}=A_{\nu;\mu}-A_{\mu;\nu}$.

The Einstein equation becomes as follows:
\begin{eqnarray}\label{eq:Einstein}
G_{\mu\nu} = 8 \pi \left( T^{\mathrm{BD}}_{\mu\nu} + \Phi^{\beta-1} T^{\mathrm{M}}_{\mu\nu} \right) \equiv 8 \pi T_{\mu\nu},
\end{eqnarray}
where the Brans-Dicke part of the energy-momentum tensors are
\begin{eqnarray}\label{eq:T_BD}
T^{\mathrm{BD}}_{\mu\nu} = \frac{1}{8\pi \Phi} \left(-g_{\mu\nu}\Phi_{;\rho \sigma}g^{\rho\sigma}+\Phi_{;\mu\nu}\right)
+ \frac{\omega}{8\pi \Phi^{2}} \left(\Phi_{;\mu}\Phi_{;\nu}-\frac{1}{2}g_{\mu\nu}\Phi_{;\rho}\Phi_{;\sigma}g^{\rho\sigma}\right)
\end{eqnarray}
and the matter part of the energy-momentum tensors are
\begin{eqnarray}\label{eq:matter}
T^{\mathrm{M}}_{\mu\nu} = T^{\mathrm{C}}_{\mu\nu} + \langle \hat{T}^{\mathrm{H}}_{\mu\nu} \rangle,
\end{eqnarray}
where
\begin{eqnarray}\label{eq:T_C}
T^{\mathrm{C}}_{\mu\nu} &=& \frac{1}{2}\left(\phi_{;\mu}\overline{\phi}_{;\nu}+\overline{\phi}_{;\mu}\phi_{;\nu}\right)
\nonumber \\
&& {}+\frac{1}{2}\left(-\phi_{;\mu}ieA_{\nu}\overline{\phi}+\overline{\phi}_{;\nu}ieA_{\mu}\phi+\overline{\phi}_{;\mu}ieA_{\nu}\phi-\phi_{;\nu}ieA_{\mu}\overline{\phi}\right)
\nonumber \\
&& {}+\frac{1}{4\pi}F_{\mu \rho}{F_{\nu}}^{\rho}+e^{2}A_{\mu}A_{\nu}\phi\overline{\phi}+\mathcal{L}^{\mathrm{EM}}g_{\mu \nu}.
\end{eqnarray}
Here, we define
\begin{eqnarray}
\mathcal{L}^{\mathrm{EM}} \equiv - \frac{1}{2}\left(\phi_{;\mu}+ieA_{\mu}\phi \right)g^{\mu\nu}\left(\overline{\phi}_{;\nu}-ieA_{\nu}\overline{\phi}\right)-\frac{1}{16\pi}F_{\mu\nu}F^{\mu\nu}
\end{eqnarray}
and $\langle \hat{T}^{\mathrm{H}}_{\mu\nu} \rangle$ is the renormalized energy-momentum tensor to include Hawking radiation, while in this paper, we will not consider this semi-classical term.

The field equations are as follows:
\begin{eqnarray}
\label{eq:Phi}0 &=& \Phi_{;\mu\nu}g^{\mu\nu}-\frac{8\pi \Phi^{\beta}}{3+2\omega} \left(T^{\mathrm{M}} - 2\beta \mathcal{L}^{\mathrm{EM}} \right) , \\
\label{eq:phi}0 &=& \phi_{;\mu\nu}g^{\mu\nu}+ieA^{\mu}\left(2\phi_{;\mu}+ieA_{\mu}\phi\right)+ieA_{\mu;\nu}g^{\mu\nu}\phi + \frac{\beta}{\Phi}\Phi_{;\mu}\left(\phi_{;\nu}+ieA_{\nu}\phi\right)g^{\mu\nu},
\\
\label{eq:A}0 &=& \frac{1}{2\pi}\left({F^{\nu}}_{\mu;\nu} + \frac{\beta}{\Phi}{F^{\nu}}_{\mu}\Phi_{;\nu} \right) -ie\phi\left(\overline{\phi}_{;\mu}-ieA_{\mu}\overline{\phi}\right)+ie\overline{\phi}\left(\phi_{;\mu}+ieA_{\mu}\phi\right),
\end{eqnarray}
where
\begin{eqnarray}\label{eq:T}
T^{\mathrm{M}} \equiv {T^{\mathrm{M}}}^{\mu}_{\;\mu}.
\end{eqnarray}

\subsection{Implementation to double-null formalism}

We use the double-null coordinates
\begin{eqnarray}\label{eq:doublenull}
ds^{2} = -\alpha^{2}(u,v) du dv + r^{2}(u,v) d\Omega^{2},
\end{eqnarray}
assuming spherical symmetry, where $u$ is the retarded time, $v$ is the advanced time, $d\Omega^{2} = d\theta^{2} + \sin^{2} \theta d\varphi^{2}$, where $\theta$ and $\varphi$ are angular coordinates.

We introduce auxiliary variables, following the notation of \cite{Yeom3}: The metric function $\alpha$, the radial function $r$, the Brans-Dicke field $\Phi$ and a complex scalar field $s \equiv \sqrt{4\pi} \phi$ and define
\begin{eqnarray}\label{eq:conventions}
h \equiv \frac{\alpha_{,u}}{\alpha},\quad d \equiv \frac{\alpha_{,v}}{\alpha},\quad f \equiv r_{,u},\quad g \equiv r_{,v},\quad W \equiv \Phi_{,u},\quad Z \equiv \Phi_{,v}, \quad w \equiv s_{,u},\quad z \equiv s_{,v}.
\end{eqnarray}

The Einstein tensor is then given as follows:
\begin{eqnarray}
\label{eq:Guu}G_{uu} &=& -\frac{2}{r} \left(f_{,u}-2fh \right),\\
\label{eq:Guv}G_{uv} &=& \frac{1}{2r^{2}} \left( 4 rf_{,v} + \alpha^{2} + 4fg \right),\\
\label{eq:Gvv}G_{vv} &=& -\frac{2}{r} \left(g_{,v}-2gd \right),\\
\label{eq:Gthth}G_{\theta\theta} &=& -4\frac{r^{2}}{\alpha^{2}} \left(d_{,u}+\frac{f_{,v}}{r}\right).
\end{eqnarray}
Also, we can obtain the energy-momentum tensors for the Brans-Dicke field part and the scalar field part:
\begin{eqnarray}
\label{eq:TBDuu}T^{\mathrm{BD}}_{uu} &=& \frac{1}{8 \pi \Phi} (W_{,u} - 2hW) + \frac{\omega}{8 \pi \Phi^{2}} W^{2},\\
\label{eq:TBDuv}T^{\mathrm{BD}}_{uv} &=& - \frac{Z_{,u}}{8 \pi \Phi} - \frac{gW+fZ}{4 \pi r \Phi},\\
\label{eq:TBDvv}T^{\mathrm{BD}}_{vv} &=& \frac{1}{8 \pi \Phi} (Z_{,v} - 2dZ) + \frac{\omega}{8 \pi \Phi^{2}} Z^{2},\\
\label{eq:TBDthth}T^{\mathrm{BD}}_{\theta\theta} &=& \frac{r^{2}}{2 \pi \alpha^{2} \Phi} Z_{,u} + \frac{r}{4 \pi \alpha^{2} \Phi} (gW+fZ) + \frac{\omega}{4\pi \Phi^{2}} \frac{r^{2}}{\alpha^{2}}WZ,
\end{eqnarray}
\begin{eqnarray}
\label{eq:TSuu}T^{\mathrm{C}}_{uu} &=& \frac{1}{4\pi} \left[ w\overline{w} + iea(\overline{w}s-w\overline{s}) +e^{2}a^{2}s\overline{s} \right],\\
\label{eq:TSuv}T^{\mathrm{C}}_{uv} &=& \frac{{(a_{,v})}^{2}}{4\pi\alpha^{2}},\\
\label{eq:TSvv}T^{\mathrm{C}}_{vv} &=& \frac{1}{4\pi} z\overline{z},\\
\label{eq:TSthth}T^{\mathrm{C}}_{\theta\theta} &=& \frac{r^{2}}{4\pi\alpha^{2}} \left[ (w\overline{z}+z\overline{w}) + iea(\overline{z}s-z\overline{s})+\frac{2{(a_{,v})}^{2}}{\alpha^{2}} \right],
\end{eqnarray}
where we will define $q(u,v) \equiv 2r^{2} a_{,v}/\alpha^{2}$ as the charge function.

To implement the double-null formalism into our numerical scheme, it is convenient to represent all equations as first-order, mixed-derivative differential equations. Note that
\begin{eqnarray}\label{eq:T_trace}
T^{\mathrm{M}} &=& - \frac{4}{\alpha^{2}}T^{\mathrm{M}}_{uv} + \frac{2}{r^{2}}T^{\mathrm{M}}_{\theta \theta},\\
\mathcal{L}^{\mathrm{EM}} &=& \frac{1}{4\pi\alpha^{2}} \left( w\bar{z} + z\bar{w} \right) + \frac{iea}{4\pi\alpha^{2}} \left( \bar{z}s - z\bar{s}\right) + \frac{{a_{,v}}^{2}}{2\pi \alpha^{4}}.
\end{eqnarray}
The Einstein equations for $\alpha_{,uv}$, $r_{,uv}$, and the field equation for $\Phi$ are then coupled as follows (we define $\widetilde{X}\equiv\Phi^{\beta}X$ for any quantity $X$):
\begin{eqnarray}\label{eq:coupled}
\left( \begin{array}{ccc}
1 & \frac{1}{r} & \frac{1}{\Phi} \\
0 & 1 & \frac{r}{2\Phi} \\
0 & 0 & r
\end{array} \right)
\left( \begin{array}{c}
d_{,u} \\
f_{,v} \\
Z_{,u}
\end{array} \right)
= \left( \begin{array}{c}
\mathfrak{A} \\
\mathfrak{B} \\
\mathfrak{C}
\end{array} \right),
\end{eqnarray}
where
\begin{eqnarray}
\label{eq:A}\mathfrak{A} &\equiv& -\frac{2\pi \alpha^{2}}{r^{2}\Phi}\widetilde{T}^{\mathrm{C}}_{\theta\theta} - \frac{1}{2r}\frac{1}{\Phi}(gW+fZ) -\frac{\omega}{2\Phi^{2}}WZ, \\
\label{eq:B}\mathfrak{B} &\equiv& - \frac{\alpha^{2}}{4r} - \frac{fg}{r} + \frac{4 \pi r}{\Phi}\widetilde{T}^{\mathrm{C}}_{uv} - \frac{1}{\Phi}(gW+fZ), \\
\label{eq:C}\mathfrak{C} &\equiv& - fZ - gW - \frac{2\pi r \alpha^{2}}{3+2\omega} \left( \widetilde{T}^{\mathrm{C}} - 2 \beta {\widetilde{\mathcal{L}}}^{\mathrm{EM}} \right).
\end{eqnarray}

After solving these coupled equations, we can write all equations:
\begin{eqnarray}\label{eq:solved}
\left( \begin{array}{c}
d_{,u}=h_{,v} \\
r_{,uv}  \\
\Phi_{,uv} 
\end{array} \right)
= \frac{1}{r^{2}} \left( \begin{array}{ccc}
r^{2} & -r & -\frac{r}{2\Phi} \\
0 & r^{2} & -\frac{r^{2}}{2\Phi} \\
0 & 0 & r
\end{array} \right)
\left( \begin{array}{c}
\mathfrak{A} \\
\mathfrak{B} \\
\mathfrak{C}
\end{array} \right),
\end{eqnarray}
\begin{eqnarray}
\label{eq:E1}r_{,uu} &=& 2fh - \frac{r}{2 \Phi} (W_{,u}-2hW) - \frac{r \omega}{2 \Phi^{2}} W^{2} - \frac{4 \pi r}{\Phi} {\widetilde{T}}^{\mathrm{M}}_{uu},\\
\label{eq:E2}r_{,vv} &=& 2gd - \frac{r}{2 \Phi} (Z_{,v}-2dZ) - \frac{r \omega}{2 \Phi^{2}} Z^{2} - \frac{4 \pi r}{\Phi} {\widetilde{T}}^{\mathrm{M}}_{vv},
\end{eqnarray}
including the field equations
\begin{eqnarray} \label{eq:fieldeqns}
\label{eq:a1}a_{,v} &=& \frac{\alpha ^{2} q}{2 r^{2}}, \\
\label{eq:q1}q_{,v} &=& -\frac{ier^{2}}{2} (\overline{s}z-s\overline{z}) - \beta q \frac{Z}{\Phi}, \\
a_{,vv} &=& \frac{\alpha^{2}}{r^{2}} \left( d - \frac{g}{r} \right)q - \frac{ie\alpha^{2}}{4} \left( z\overline{s}-s\overline{z}\right) - \beta q \frac{\alpha^{2} Z}{2r^{2}\Phi},\\
q_{,u} &=& \frac{ier^{2}}{2} (\overline{s}w-s\overline{w}) - r^{2}e^{2}a s \overline{s} - \beta q \frac{W}{\Phi}, \\
\label{eq:a2}a_{,uv} &=& \frac{\alpha^{2}}{r^{2}} \left( h - \frac{f}{r} \right)q + \frac{ie\alpha^{2}}{4} \left( w\overline{s}-s\overline{w}\right) - \frac{\alpha^{2}}{2}e^{2}as\overline{s} - \beta q \frac{\alpha^{2} W}{2r^{2}\Phi},\\
\label{eq:s}s_{,uv} &=& - \frac{fz}{r} - \frac{gw}{r} - \frac{iearz}{r} - \frac{ieags}{r} - \frac{ie}{4r^{2}}\alpha^{2}qs - \frac{\beta}{2\Phi} \left( Wz +Zw + ies a Z \right),
\end{eqnarray} 
where Equations~(\ref{eq:solved}) and (\ref{eq:s}) are evolution equations for $\alpha_{,uv}$, $r_{,uv}$, $\Phi_{,uv}$ and $s_{,uv}$, which can be solved with the same integration scheme used in the authors' previous papers \cite{authors,Hansen:2013vha}. The gauge field $a$ can be evolved using the same numerical scheme via Equation~(\ref{eq:a2}) or we can evolve $a$ and $q$ via Equations~(\ref{eq:a1}) and (\ref{eq:q1}). We have found the latter method to be more robust for this study and have used it for our integration via a standard 4th order Runge-Kutta scheme. Consistency and convergence tests for our numerical implementation of the present set of equations are presented in Appendix~B.

\subsection{\label{sec:ini}Initial conditions and free parameters}

For our simulations, we need to specify initial conditions for the dynamic variables ($\alpha, r, \Phi, s, a$) on the initial $u=u_{\mathrm{i}}$ and $v=v_{\mathrm{i}}$ null surfaces. For all simulations in this paper, we choose $u_{\mathrm{i}}=v_{\mathrm{i}}=0$ and computational domain to have size $v=[0;60]$ and $u=[0;40]$.

First off, we have the gauge freedom to choose $r$ along the initial null surfaces. Here, we choose $r(u,0)_{,u}=r_{u0}<0$ and $r(0,v)_{,v}=r_{v0}>0$ such that the radial function for an in-going observer decreases and that for an out-going observer increases. In addition,　we assume that the effective gravitational constant $G=1/\Phi$ is asymptotically unity, hence we can set $\Phi(u,0)=\Phi(0,v)=1$. 

\paragraph{In-going null direction}

We use an in-going shell-shaped scalar field such that its interior is not affected by the shell. Thus, we can simply choose $s(u,0)=0$, $\alpha(u,0)=1$, $q(u,0)=0$ and $a(u,0)=0$. Then, the Misner-Sharpe mass function
\begin{eqnarray}
m(u,v) \equiv \frac{r}{2}\left( 1+\frac{q^{2}}{r^{2}}+4\frac{r_{,u}r_{,v}}{\alpha^{2}} \right)
\end{eqnarray}
should vanish at $u=v=0$. To satisfy this, it is convenient to choose $r_{,u}(u,0)=-1/2$, $r_{,v}(0,v)=1/2$. We also choose the radial function at the initial point $u=v=0$ to be $r(0,0)=r_0=20$, thereby fully determining the radial function along the initial null segments. These initial conditions are in agreement with the constraint equations (Equation~(\ref{eq:E1})) and thus completes the assignment of initial conditions along the in-going null segment.

\paragraph{Out-going null direction}

We can choose an arbitrary function for $s(0,v)$ to induce a collapsing, in-going pulse. In this paper, we use 
\begin{eqnarray} \label{s_initial}
s(0,v)= A \sin^{4} \left( \pi \frac{v-v_{\mathrm{i}}}{v_{\mathrm{f}}-v_{\mathrm{i}}} \right) \left[ \cos \left( 2 \pi \frac{v-v_{\mathrm{i}}}{v_{\mathrm{f}}-v_{\mathrm{i}}} \right) + i \cos \left( 2 \pi \frac{v-v_{\mathrm{i}}}{v_{\mathrm{f}}-v_{\mathrm{i}}} -  \pi \delta \right)\right]
\end{eqnarray}
for $v_{\mathrm{i}}\leq v \leq v_{\mathrm{f}}$ and otherwise $s(0,v)=0$. Next, we can integrate constraint equation Equation~(\ref{eq:E2}) to determine $\alpha(0,v)$ on the $u=0$ surface, simultaneously with integrating Equations~(\ref{eq:a1}) and (\ref{eq:q1}) for determining $q(0,v)$ and $a(0,v)$. 

We have now fully specified the five dynamic variables ($\alpha, r, \Phi, s, a$) along the both in- and out-going null segments in a manner consistent with the constraint equations (Equations~(\ref{eq:E1}) and (\ref{eq:E2})). This finishes the basic assignment of initial conditions (but cf. with paragraph below).

\paragraph{Free parameters}

For the scalar field pulse, we choose $v_{\mathrm{f}}=20$, $A = 0.15$ and $\delta = 0.5$ (to include maximum charge), leaving just three free parameters $(\omega, \beta, e)$, where $\omega$ is the Brans-Dicke coupling parameter, $\beta$ is the coupling between the matter sector and the Brans-Dicke field and $e$ is the gauge coupling. To summarize, we vary as follows:
\begin{description}
\item[-- $\omega$]: $0$ ($f(R)$ limit), $-1$ (dilaton limit), $-1.4$ (braneworld limit), $-1.6$ (ghost limit),
\item[-- $\beta$]: $0$ (Type IIA), $0.5$ (Type I), $1$ (Heterotic),
\item[-- $e$]: $0$ (neutral black hole), $0.3$ (charged black hole).
\end{description}

\begin{figure}
\begin{center}
\includegraphics[scale=0.17]{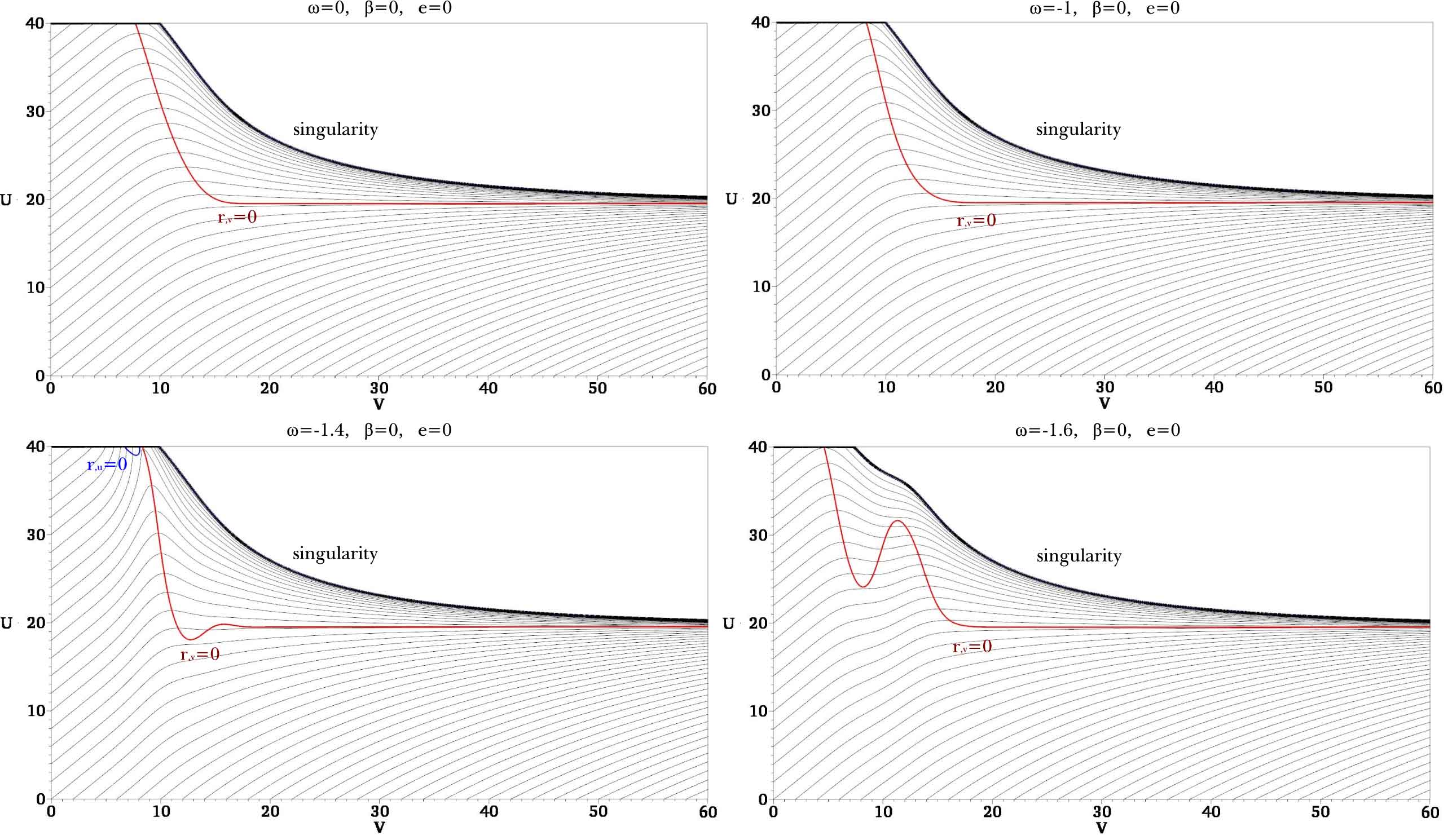}
\caption{\label{fig:neutral_beta0} Causal structure of neutral black holes with $\beta=0$. Thin black curves are contours of a constant $r$; the difference between contours are $1$. Red curve is the apparent horizon $r_{,v}=0$, blue curve is the horizon $r_{,u}=0$, and thick black curve is the singularity $r=0$. All subsequent causal structure plots has similar legend.}
\end{center}
\end{figure}

\begin{figure}
\begin{center}
\includegraphics[scale=0.17]{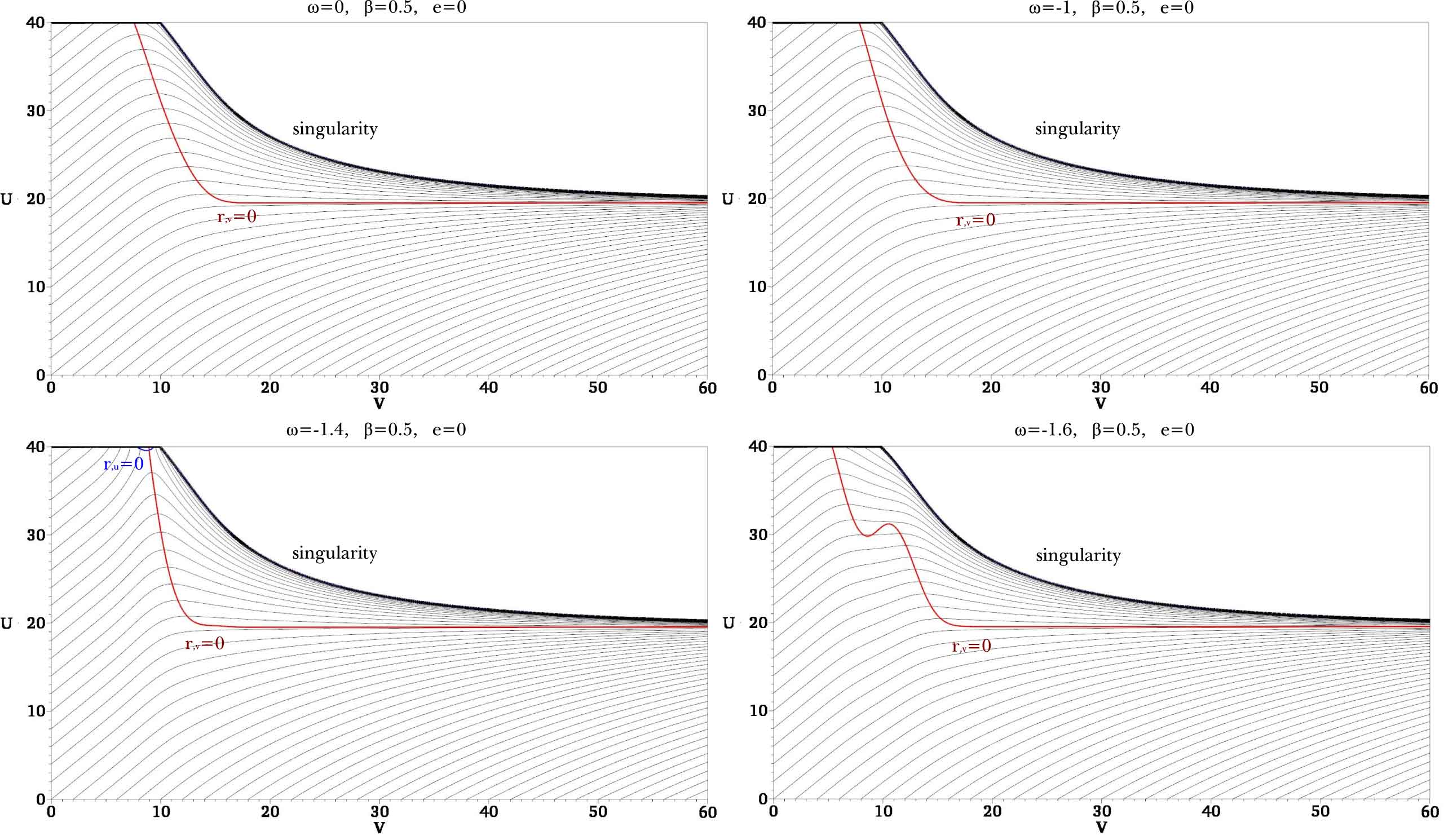}
\caption{\label{fig:neutral_beta05} Causal structure of neutral black holes with $\beta=0.5$.}
\end{center}
\end{figure}

\begin{figure}
\begin{center}
\includegraphics[scale=0.17]{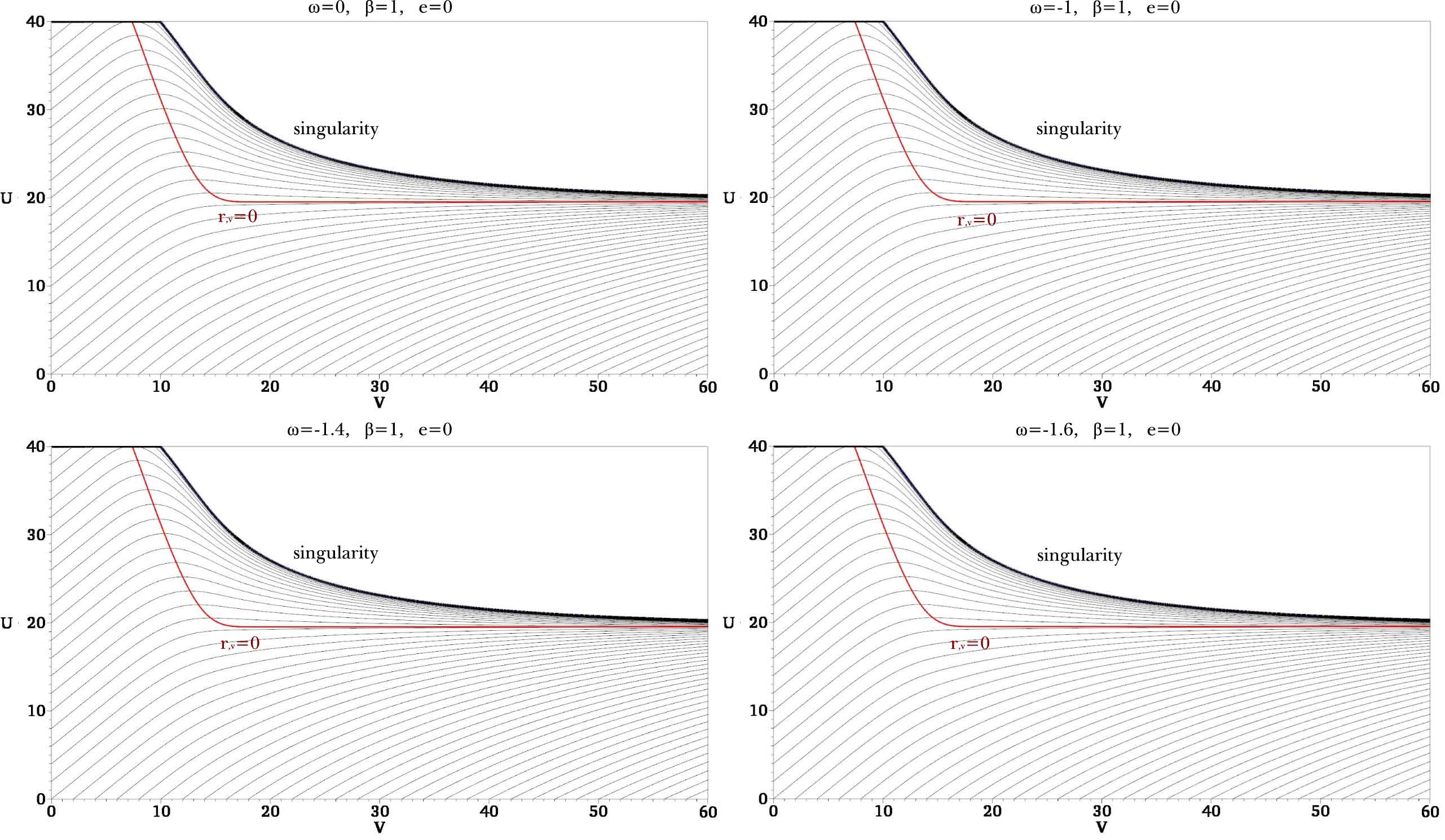}
\caption{\label{fig:neutral_beta1} Causal structure of neutral black holes with $\beta=1$.}
\end{center}
\end{figure}

\begin{figure}
\begin{center}
\includegraphics[scale=0.7]{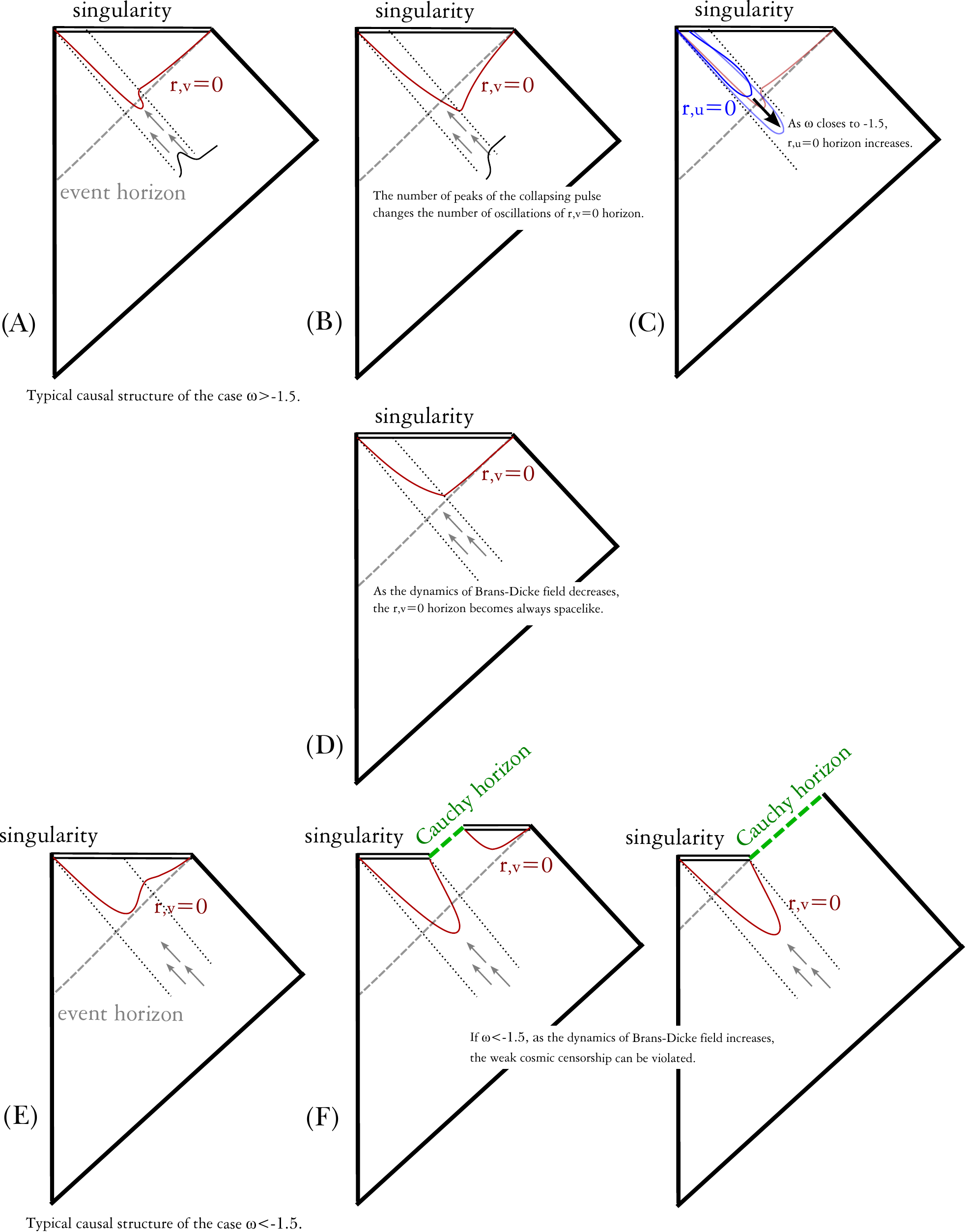}
\caption{\label{fig:classification} Classification of causal structures for neutral black holes.}
\end{center}
\end{figure}

\section{\label{sec:dyn}Causal structures and dynamics of Brans-Dicke field}

In this section, we report the possible causal structures and responses of the Brans-Dicke field for neutral and charged black holes of various string-inspired models obtained from numerical calculations by using the double-null formalism.

\subsection{Causal structures}

\subsubsection{Neutral black holes}

In this paper, we wish to study charged black holes. But first, let us briefly summarize the case of neutral black holes; Figure~\ref{fig:neutral_beta0} shows the causal structures, plotted are lines of constant r, apparent horizons and $r=0$ singularities\footnote{We remind the reader of the ability of double-null codes to numerically integrate spacetimes with spacelike singularities (without crashing the code) all the way up to the singularity, see references in appendinx B.} of neutral black holes for $\beta=0$, Figure~\ref{fig:neutral_beta05} shows neutral black holes for $\beta=0.5$ and Figure~\ref{fig:neutral_beta1} shows neutral black holes for $\beta=1$,  all figures are obtained by our numerical code. 

Let us briefly look at the Brans-Dicke field equation in the neutral limit:
\begin{eqnarray}
0 = \Phi_{;\mu\nu}g^{\mu\nu}-\frac{8\pi \Phi^{\beta}}{3+2\omega} \left(T^{\mathrm{M}} - 2\beta \mathcal{L}^{\mathrm{EM}} \right),\nonumber
\end{eqnarray}
where
\begin{eqnarray}
T^{\mathrm{M}} - 2\beta \mathcal{L}^{\mathrm{EM}} =  \frac{1-\beta}{2\pi \alpha^{2}} \left(w\bar{z} + z\bar{w}\right).
\end{eqnarray}
Therefore, if $\omega > -3/2$, then since the term $T^{\mathrm{M}} - 2\beta \mathcal{L}^{\mathrm{EM}}$ is positive, it pushes the Brans-Dicke field to the strong coupling limit ($\Phi < 1$) and vice versa. If $\omega < -3/2$ then it behaves oppositely. The $\beta = 1$ case is unique and in this case, there is no back-reactions to the Brans-Dicke field and hence there is no dependence on $\omega$ (the four figures in Figure~\ref{fig:neutral_beta1} are the same, bar numerical errors).

\begin{figure}
\begin{center}
\includegraphics[scale=0.17]{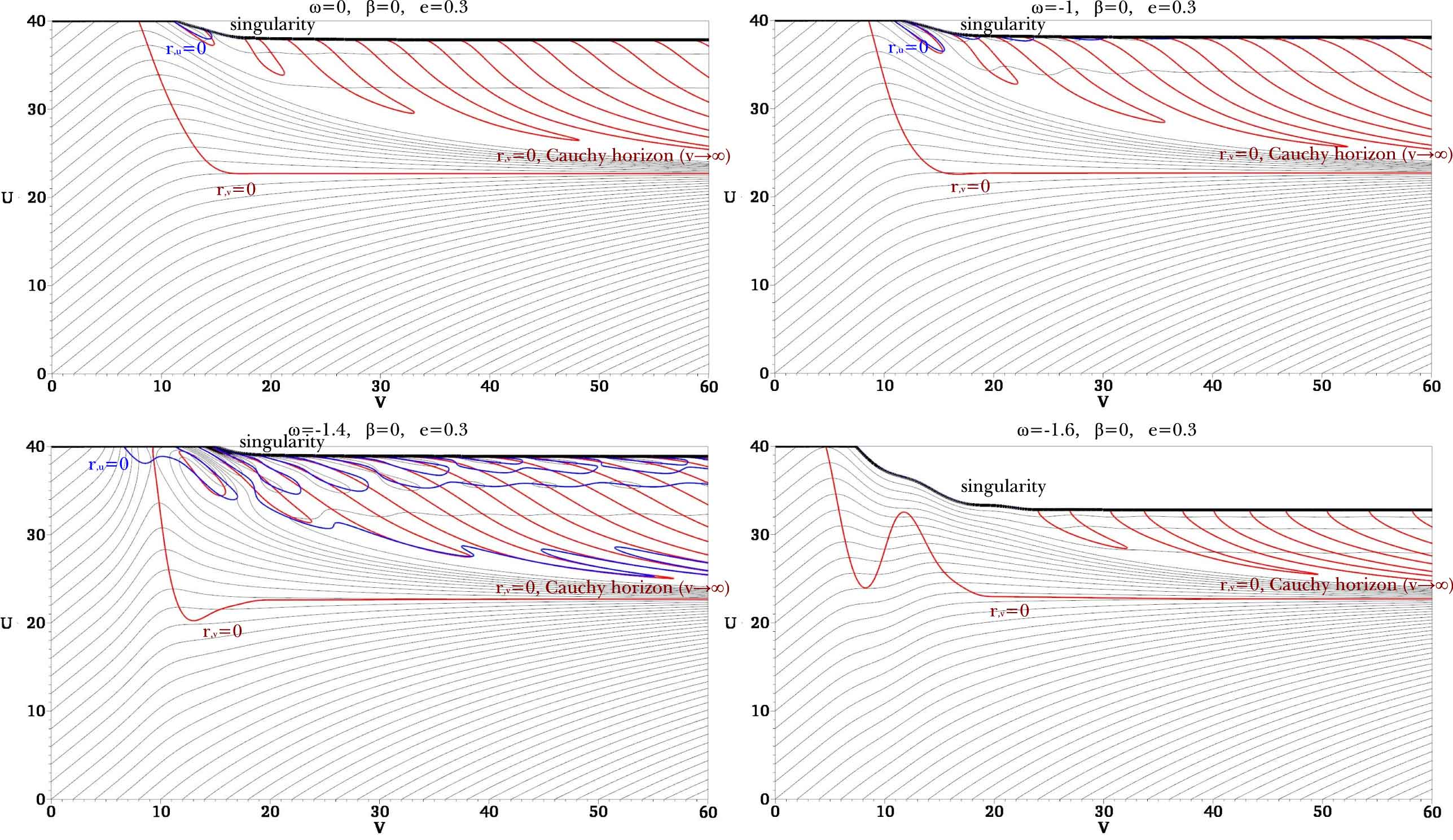}
\caption{\label{fig:charge_beta0} Causal structure for charged black holes with $\beta=0$.}
\end{center}
\end{figure}

As we observed in a previous paper \cite{Yeom2}, if $\omega > -3/2$, then during the gravitational collapse, the Brans-Dicke field is pushed to $\Phi < 1$ limit and after the collapse, the pushed field is repulsed and approaches the stationary limit. On the other hand, if $\omega < -3/2$, then Brans-Dicke is pushed to $\Phi > 1$ limit during gravitational collapses and approaches the stationary limit. Because of this behaviour, there appears not only the apparent horizon $r_{,v}=0$, but also $r_{,u}=0$ for $\omega > -3/2$. In addition, the apparent horizon can show strong oscillations for $\omega < -3/2$ and perhaps it can create a Cauchy horizon as a naked singularity. As $1-\beta$ decreases, the back-reaction decreases. This is clear, if we observe the location of $r_{,u}=0$ horizons or the oscillations of the apparent horizon $r_{,v}=0$ and compare with the $\beta = 0$ case and the $\beta = 0.5$; they tend to disappear, where these phenomena are the typical behaviors which are due to the Brans-Dicke field.

We summarize the possible charge-less causal structures, inferred from our simulations, in Figure~\ref{fig:classification}. For $\omega > -3/2$, the typical shape is (A). The oscillations of the apparent horizon $r_{,v}=0$ depends on the number of peaks of the initial collapsing field (see (B) and Figure~\ref{fig:Tvv_beta0} in Appendix~C, and compare with Figure~5 of \cite{Yeom2}). This is one difference between \cite{Yeom2} and the current paper. The location of $r_{,u} = 0$ horizon appears and increases as the dynamics of the Brans-Dicke field increases, e.g., as $\omega$ approaches $-3/2$ or $1-\beta$ increases, as denoted in (C). For $\omega < -3/2$, the typical shape is (E), while the apparent horizon and the singularity can meet as was observed in \cite{Yeom2} and the causal structure can be (F). For all cases, when the dynamics of the Brans-Dicke field is suppressed, the causal structure approaches (D) and follows that of Einstein gravity.

\subsubsection{Charged black holes}

Now we turn on the gauge coupling parameter $e=0.3$. Note that the case of $\delta = 0.5$ is the case with maximal asymptotic charge (for Einstein gravity, this is demonstrated in \cite{Yeom3}), since the phase between the real part and the imaginary part become maximal. However, in general it does not mean that extreme (or naked) black holes can be formed (this was also demonstrated by \cite{Yeom3}). The excessive charge will be repelled outside the horizon and the excessive charge is either repelled to future infinity or forms a stable charge cloud outside the horizon \cite{Hansen:2013vha}. Hence in terms of gravitational collapses, $\delta = 0.5$ allows the largest possible amount of charge.

\begin{figure}
\begin{center}
\includegraphics[scale=0.17]{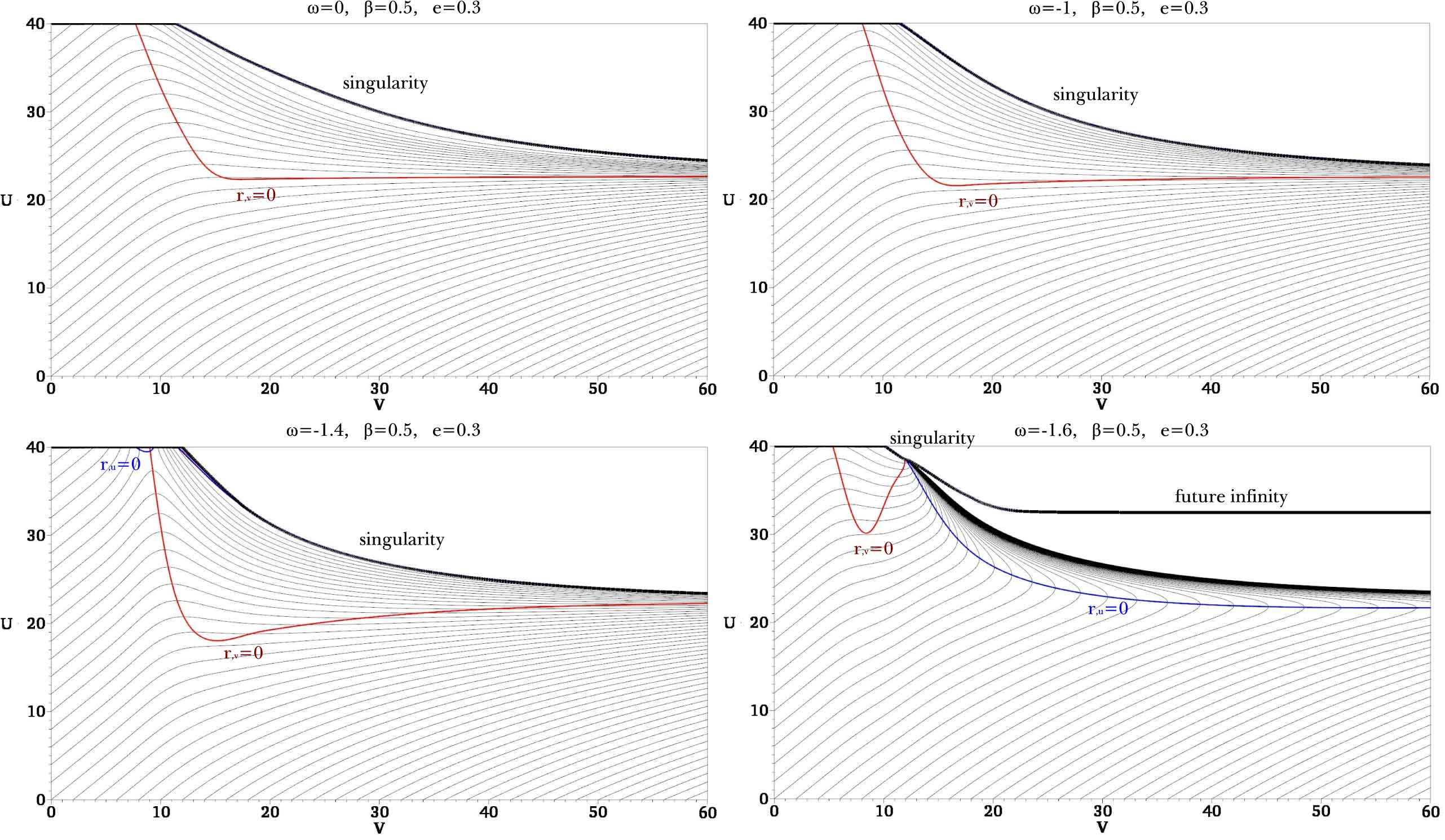}
\caption{\label{fig:charge_beta05}Causal structure for charged black holes with $\beta=0.5$.}
\end{center}
\end{figure}

\paragraph{Case $\beta=0$:} Figure~\ref{fig:charge_beta0} shows causal structures of charged black holes with $\beta = 0$. The obvious feature is the existence of a Cauchy horizon inside the outer apparent horizon. The singularity is always spacelike. However, in the limit $v \rightarrow \infty$, there will be nonzero distance (in terms of $u$-coordinate) between the outer apparent horizon and the singularity. This indicates the existence of the Cauchy horizon. These global behaviors are the same as for the case of charged black holes in Einstein gravity. In the left of Figure~\ref{fig:alpha_maintext} (as well as in Figure~\ref{fig:alpha_beta0} in Appendix~C), the metric function $\alpha$ is plotted and there it can clearly be seen that inside the horizon, $\alpha \rightarrow 0$ and which hence indicates mass inflation inside the charged black hole (see Appendix~A).

One important difference between this and the previous case, is the complicated horizon behaviors inside the apparent horizon. Such non-trival behaviours are to be expected since we have both the dynamics of the matter field as well as the Brans-Dicke field inside the horizon and these cause the observed complicated dynamical behaviour of horizons. In Einstein gravity, if we do not turn on the semi-classical effects, then there is no complicated horizon dynamics inside the horizon; But, if we turn on the semi-classical effects, then the null energy condition is violated and hence the inside horizon structure becomes complicated \cite{Yeom3}. Also, for modified gravity models such as $f(R)$ gravity (in the Jordan frame), similar behaviors has been observed \cite{Yeom5}.

\begin{figure}
\begin{center}
\includegraphics[scale=0.17]{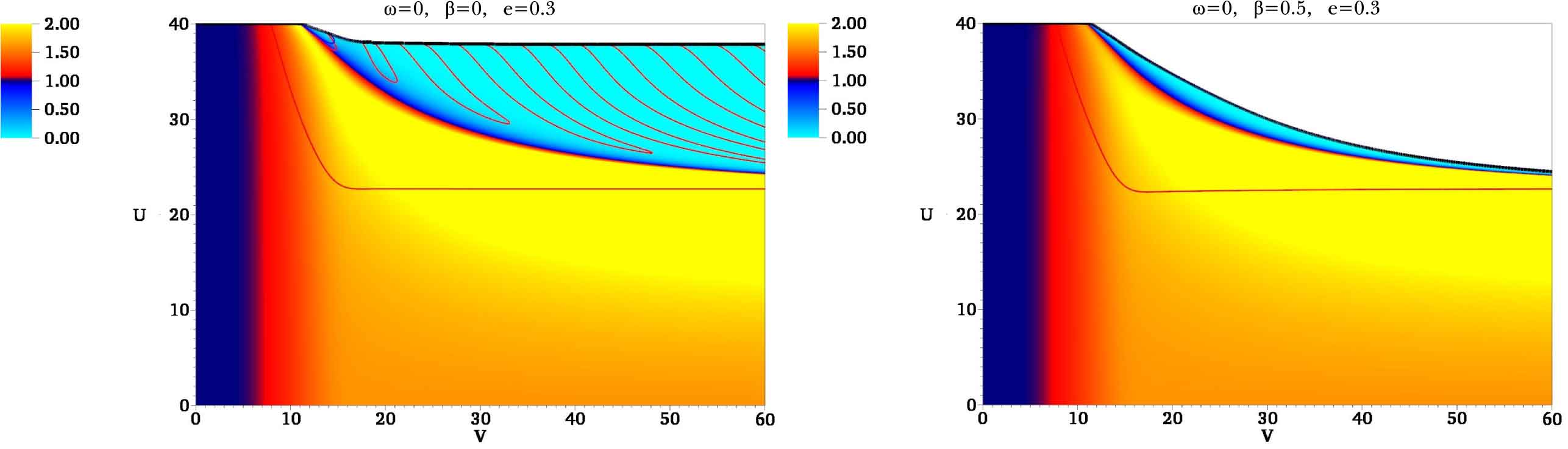}
\caption{\label{fig:alpha_maintext}Comparison of $\alpha$ for $\beta = 0$ (left) and $\beta = 0.5$ (right), with the same $\omega=0$ and $e=0.3$. The cases for this and other initial conditions are illustrated in Appendix~C.}
\end{center}
\end{figure}

\begin{figure}
\begin{center}
\includegraphics[scale=0.17]{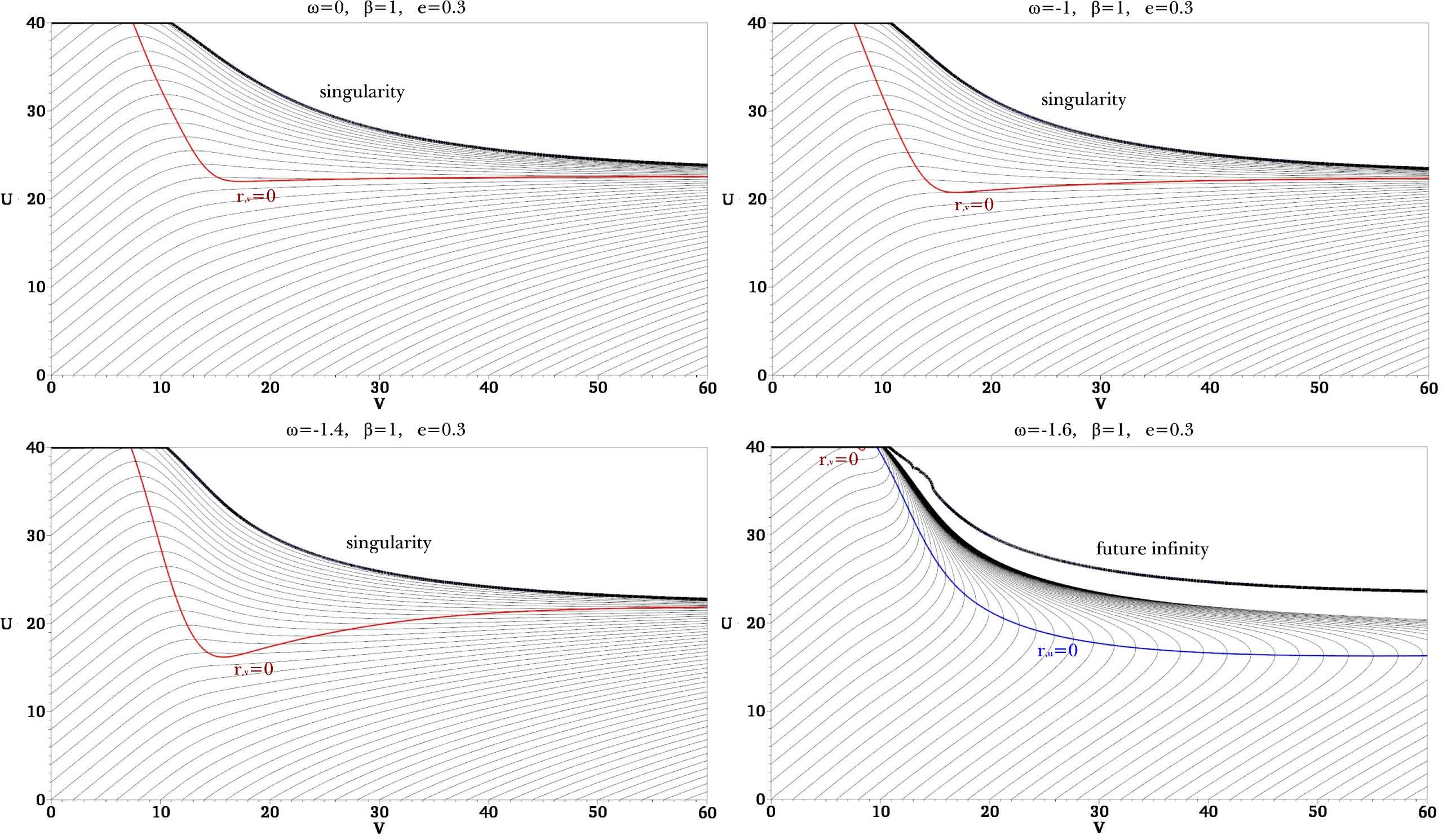}
\caption{\label{fig:charge_beta1}Causal structure for charged black holes with $\beta=1$.}
\end{center}
\end{figure}

\paragraph{Case $\beta=0.5$:} Figure~\ref{fig:charge_beta05} shows causal structure for the case of $\beta = 0.5$. Comparing this with the $\beta = 0$ case, we see to two main differences.

First, for $\omega > -3/2$, the Cauchy horizon \textit{disappears}. In other words, there is a spacelike singularity and it approaches the apparent horizon in the $v \rightarrow \infty$ limit. This was observed by \cite{Ann} (for dilaton limit in the Einstein frame) and our results (various $\omega$ in the Jordan frame) are consistent with it. This issue will be clarified when we study the response of Brans-Dicke field in the next section. Looking at the $\alpha$ function, there is the signature of mass inflation, see for example right of Figure~\ref{fig:alpha_maintext} for $e=0.3$ and $\omega > -3/2$ (as well as in Figure~\ref{fig:alpha_beta05} in Appendix~C), since there is a region where $\alpha$ approaches zero. However, this is not enough to generate a Cauchy horizon as in the $\beta = 0$ cases.

Secondly, for $\omega < -3/2$, there can be an $r_{,u} = 0$ horizon which will eventually form a \textit{cosmological horizon}. This was not observed in gravitational collapses of neutral matters. Therefore, we can state that if $\omega < -3/2$, then a \textit{charged} gravitational collapse can induce inflation and a baby universe. If we interpret the $r_{,u} = 0$ horizon as a throat of a (one-way traversable) wormhole, then we can make a one-way traversable wormhole using a charged gravitational collapse. Of course, we know that if we consider the limit $\omega < -3/2$, then the Brans-Dicke field becomes a ghost and hence it can allow a wormhole or a baby universe \cite{Agnese:1995kd}. However, what we show is that such objects can be obtained just by \textit{gravitational collapse} of a charged matter field, i.e., in an extremely simple way while previously, complicated initial conditions were needed to obtain such a state \cite{JHY}.

\paragraph{Case $\beta=1$:} The causal structure of this case, shown in figure~\ref{fig:charge_beta1}, shares many qualitative properties with the $\beta = 0.5$ case. In the neutral limit, there was no similar behaviors for $\omega < -3/2$ limit, therefore, it is more clear that the existence of the inflating region really is  comes from the effects of \textit{charge}.

Importantly, it is noted that for the case of $\omega = -1$ (dilaton gravity with Heterotic theory), there exists a spacelike singularity. There were some discussions that extreme black holes can be regular \cite{Gibbons:1987ps}. However, if the black hole is formed from a gravitational collapse, then we have to conclude that such a regular black hole state is impossible to obtain.

\begin{figure}
\begin{center}
\includegraphics[scale=0.7]{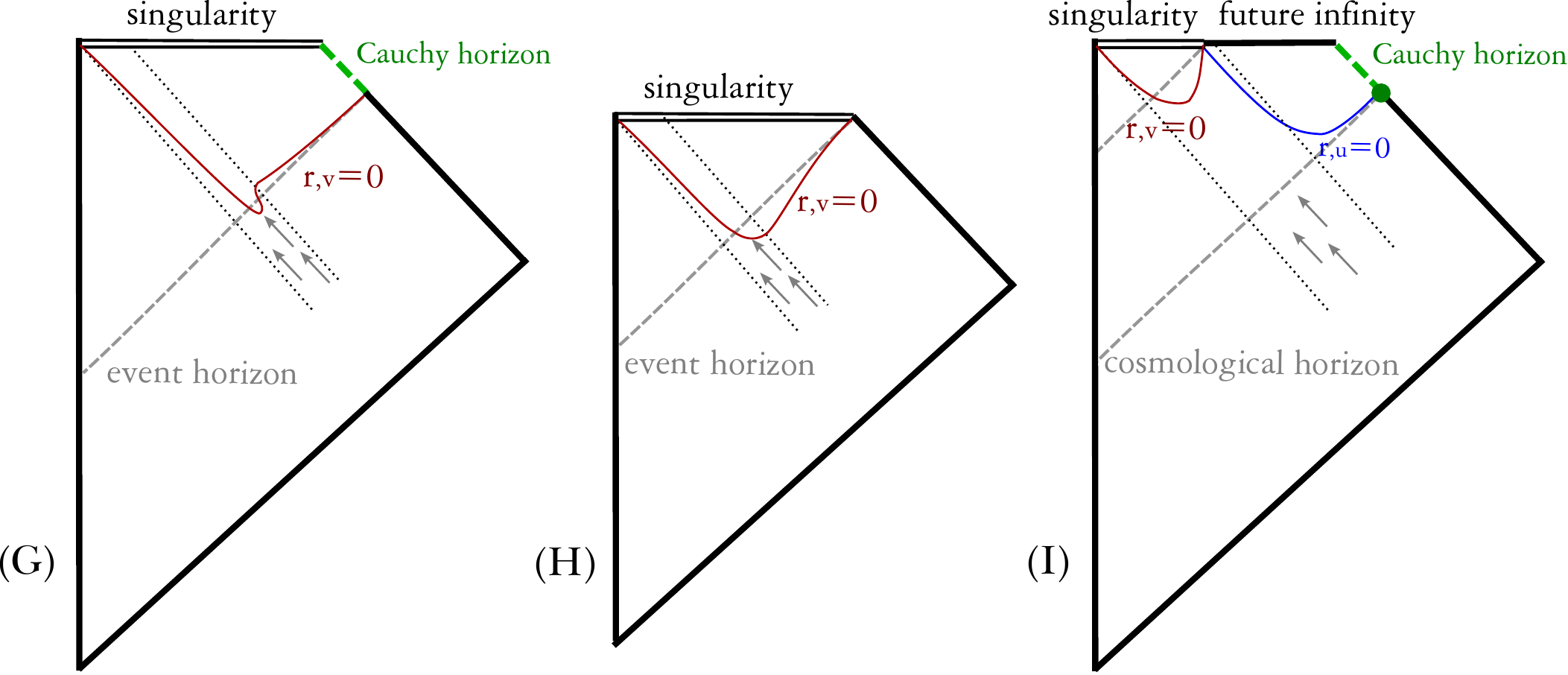}
\caption{\label{fig:causal}Causal structures for charged black holes.}
\end{center}
\end{figure}

\paragraph{Summary of causal structures} The intuitive nature of local $r_{,v}=0$ horizons or $r_{,u}=0$ horizons are the same as Figure~\ref{fig:classification}. However, when we turn on the charge, the global structures can take on three different types. (1) For $\beta = 0$, there is the Cauchy horizon ((G) of Figure~\ref{fig:causal}). (2) For $\omega > -3/2$ and $\beta =0.5$ or $1$, there is no Cauchy horizon ((H) of Figure~\ref{fig:causal}). (3) For $\omega < -3/2$ and $\beta = 0.5$ or $1$ cases, the gravitational collapse can induce inflation ((I) of Figure~\ref{fig:causal}). If the final point of $r_{,u}=0$ in the $v \rightarrow \infty$ limit (green dot in (I)) does not meet the spacelike future infinity, then there will be a Cauchy horizon in the $v \rightarrow \infty$ limit.

\begin{figure}
\begin{center}
\includegraphics[scale=0.17]{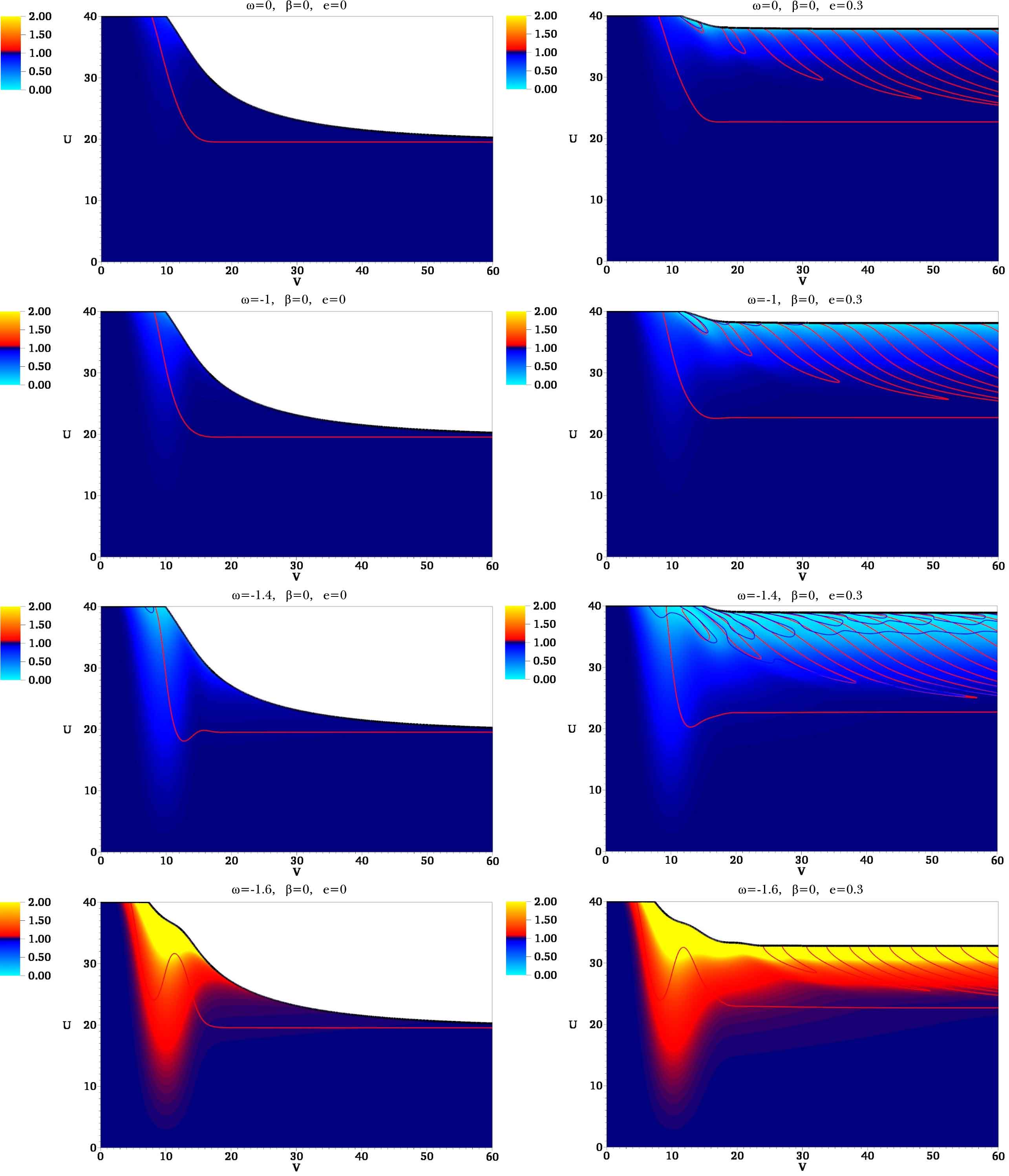}
\caption{\label{fig:BD_beta0}Plots of $\Phi$ for $\beta=0$.}
\end{center}
\end{figure}

\subsection{Response of the Brans-Dicke field}

The Brans-Dicke field equation is
\begin{eqnarray}
0 = \Phi_{;\mu\nu}g^{\mu\nu}-\frac{8\pi \Phi^{\beta}}{3+2\omega} \left(T^{\mathrm{M}} - 2\beta \mathcal{L}^{\mathrm{EM}} \right),\nonumber
\end{eqnarray}
where
\begin{eqnarray}
T^{\mathrm{M}} - 2\beta \mathcal{L}^{\mathrm{EM}} = - \beta \frac{q^{2}}{4\pi r^{4}} + \frac{1-\beta}{2\pi \alpha^{2}} \left( \left(w\bar{z} + z\bar{w}\right) + iea \left( \bar{z}s-z\bar{s} \right) \right).
\end{eqnarray}
Note that we can approximately write this as follows:
\begin{eqnarray}
0 = \Phi_{;\mu\nu}g^{\mu\nu}-\frac{8\pi \Phi^{\beta}}{3+2\omega} \left( - \beta \times \left(\mathrm{charge\;\; terms} \right) + \left( 1 - \beta \right) \times \left(\mathrm{kinetic\;\; terms} \right) \right).\nonumber
\end{eqnarray}
The kinetic terms mainly contribute \textit{during} the gravitational collapse (i.e. in our simulations for $v < v_{\mathrm{f}} = 20$), while the charge terms will contribute \textit{after} the gravitational collapse (i.e. in our simulations for $v > v_{\mathrm{f}} = 20$) and \textit{outside} the apparent horizon. Outside the black hole, the charge term will remain since the charge is a conserved quantity, while the kinetic term will disappear as the gravitational collapse ends.

If we keep this in mind, then for the $\beta = 0$ case, there will be essentially no effects after the gravitational collapse and the outside the horizon. In other words, the Brans-Dicke hair will decay as time goes on and approach a stationary limit, Figure~\ref{fig:BD_beta0} shows such behaviors. On the other hand, in the $\beta=1$ limit, there will be almost no back-reaction to the Brans-Dicke field during the gravitational collapse; Only after the gravitational collapse has finished will the Brans-Dicke field be affected. Figure~\ref{fig:BD_beta1} shows such behaviors very clearly.

\begin{figure}
\begin{center}
\includegraphics[scale=0.17]{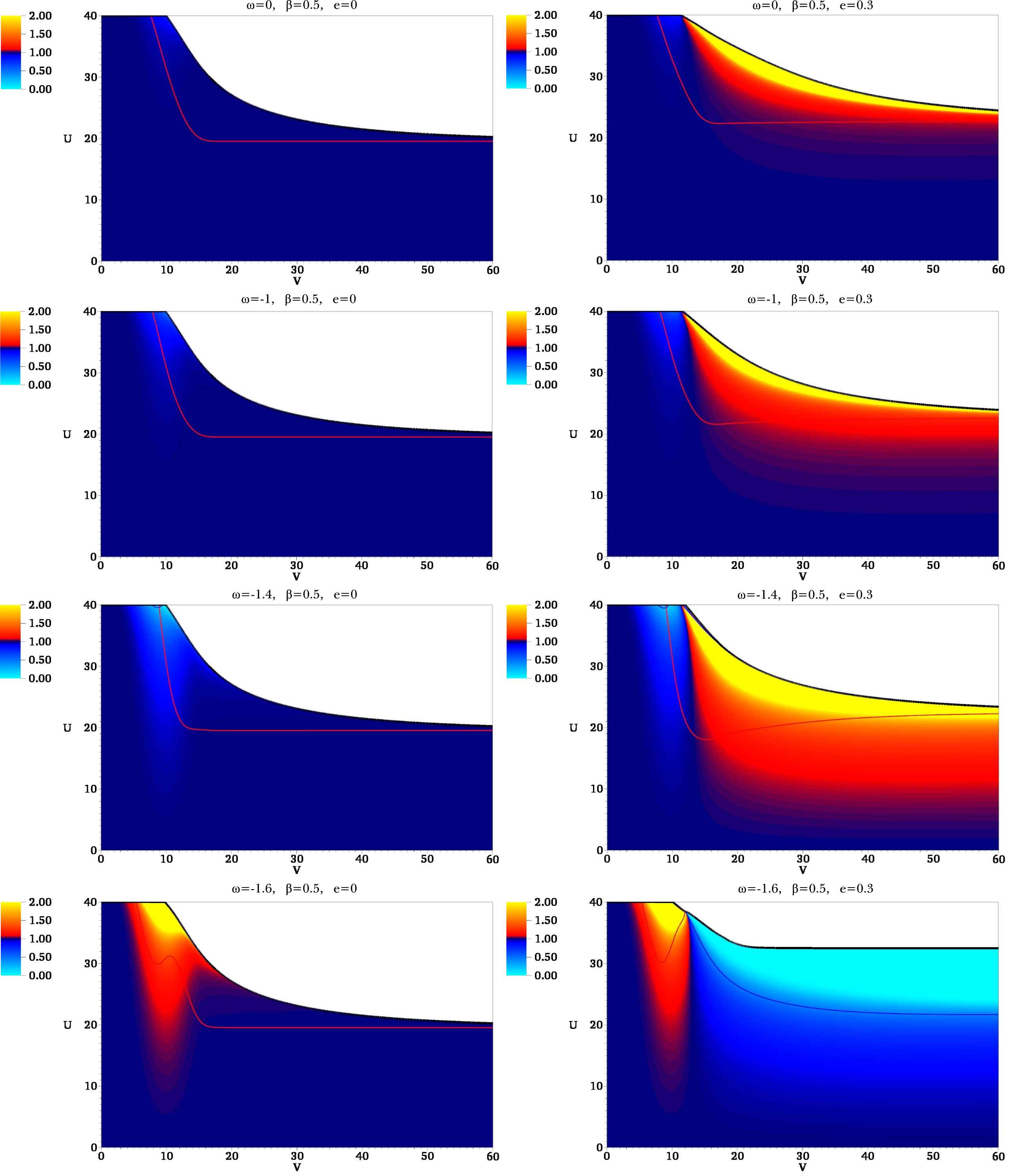}
\caption{\label{fig:BD_beta05}Plots showing the Brans-Dicke field $\Phi$ for $\beta=0.5$.}
\end{center}
\end{figure}

\begin{figure}
\begin{center}
\includegraphics[scale=0.17]{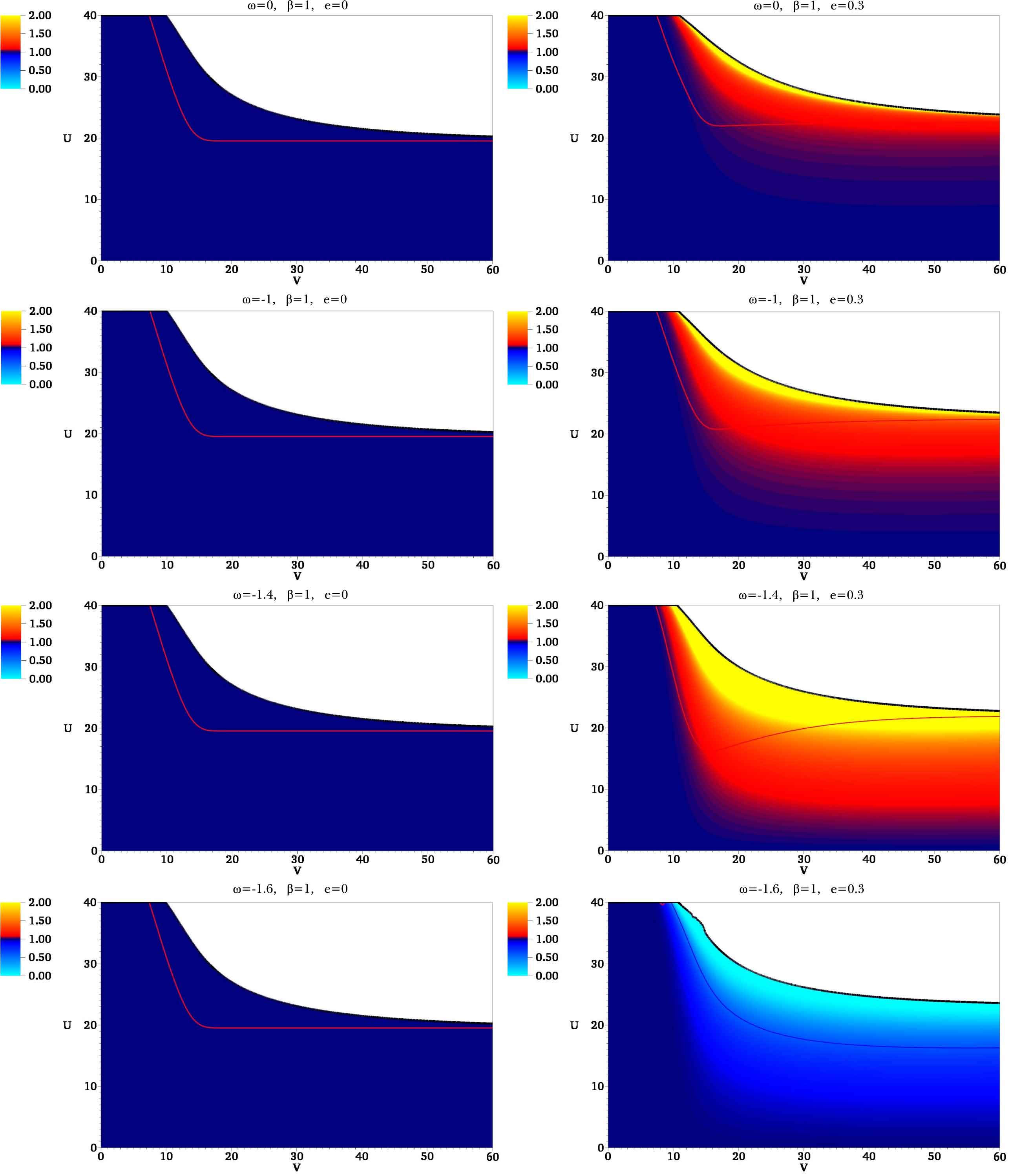}
\caption{\label{fig:BD_beta1}Plots showing the Brans-Dicke field $\Phi$ for $\beta=1$.}
\end{center}
\end{figure}

If $0 < \beta < 1$, then one can see both effects simultaneously, such behaviour is shown in figure~\ref{fig:BD_beta05}; One can see the contrast between during ($0 < v <20$) and after ($v >20$) the gravitational collapse, which is different from the cases of $\beta = 0$ and $\beta = 1$.

Now let us think about the directions of the Brans-Dicke fields, whether it becomes greater or less than one. For $\omega > -3/2$, if the term $T^{\mathrm{M}} - 2\beta \mathcal{L}^{\mathrm{EM}}$ is positive, it will push the Brans-Dicke field to the strong coupling limit ($\Phi < 1$); If it is negative, then it will push to the weak coupling limit ($\Phi > 1$). To summarize:
\begin{description}
\item[--] Effects of kinetic terms: If $\beta < 1$ and $\omega > -3/2$, then the kinetic terms always push the Brans-Dicke field to the strong coupling limit.
\item[--] Effects of charge terms: If $\beta > 0$ and $\omega > -3/2$, then the charge terms always push the Brans-Dicke field to the weak coupling limit.
\end{description}
For the other cases (e.g., $\omega < -3/2$), opposite behaviors will be observed. This is consistent with the numerical results shown in Figures~\ref{fig:BD_beta0}, \ref{fig:BD_beta05} and \ref{fig:BD_beta1}. For $0 \leq \beta \leq 1$ and $\omega > -3/2$, during the gravitational collapse, $\Phi < 1$ appears (shifts to blue color) except $\beta = 1$. After the gravitational collapse, always we have $\Phi > 1$ (shifts to red color) except for $\beta = 0$. In particular we can see both phenomena at the same time for the case of $\beta=0.5$. If $\omega < -3/2$, then all behaviors are opposite.

We can summarize our results (for $\omega > -3/2$) as follows:
\begin{description}
\item[-- $\beta < 0$:] All effects push the Brans-Dicke field to the strong coupling limit. As time goes on, the Brans-Dicke hair will be formed in the strong coupling direction.
\item[-- $\beta = 0$:] The Brans-Dicke field is pushed to the strong coupling limit, but only dynamically. As time goes on, the Brans-Dicke hair will disappear in the stationary limit.
\item[-- $0 < \beta < 1$:] The Brans-Dicke field is pushed to the strong coupling limit dynamically. On the other hand, as time goes on, the Brans-Dicke hair will be formed in the weak coupling direction.
\item[-- $\beta = 1$:] The Brans-Dicke field does not affected by gravitational collapse. However, as time goes on, the Brans-Dicke hair will be formed in the weak coupling direction.
\item[-- $1 < \beta$:] All effects push the Brans-Dicke field to the weak coupling limit.
\end{description}

One final comment should be made regarding the no-hair theorem in the Einstein frame. We know that for a quite general class of models, the no-hair theorem should be satisfied in the Einstein frame, hence, there should be no scalar hair in the stationary limit. If this is the case in the Einstein frame, then there should be no hair in the Jordan frame, too \cite{Sotiriou:2011dz}. However, our model is consistent with these general discussions. After we change the Einstein frame, the matter sector Lagrangian has a non-trivial coupling between the matter field and the Brans-Dicke field: so to speak, $\mathcal{L}_{\mathrm{E}}(\phi, A_{\mu}, \Phi)$, unless $\beta = 0$. This violates the assumptions of \cite{Sotiriou:2011dz} and it explains why a non-trivial scalar hair appears after the gravitational collapse has finished. Furthermore, we can consistently observe that there is no scalar hair after the gravitational collapses for the $\beta = 0$ case.

\section{\label{sec:dis}Discussion}

In this paper, we have investigated charged black holes of string-inspired gravity models. Two parameters are important: The Brans-Dicke parameter $\omega$ and the coupling between the Brans-Dicke field and the gauge field $\beta$. This model covers lots of string-inspired models: The dilaton gravity $\omega = -1$ with Type~IIA ($\beta = 0$), Type~I ($\beta = 0.5$), Heterotic ($\beta = 1$), some braneworld inspired models $\omega > -3/2$ and $f(R)$ inspired models $\omega = 0$. In addition, we investigated the ghost limit $\omega < -3/2$. We used numerical techniques with double-null formalism to numerically solve the dynamic equations.

We have focused on the causal structures and responses of the Brans-Dicke field, especially reactions via charges. For usual non-ghost limit $\omega > -3/2$, the direction of the response of the Brans-Dicke field, whether biased to strong coupling limit or weak coupling limit, is determined by $\beta$. If $\beta = 1$, then there is no response of the Brans-Dicke field during the gravitational collapse; If $\beta = 0$, then there is no response of the Brans-Dicke field via charges after a black hole is formed. This means that after the gravitational collapse, as long as $\beta > 0$, the Brans-Dicke field will be biased to the weak coupling limit.

This is very crucial in order to understand the nature of charged black holes for $\beta > 0$. An important point is that in general such a black hole will have Brans-Dicke hair. Another important point is that the Brans-Dicke field is biased to the weak coupling limit and this will \textit{screen} the charge inside the horizon. Therefore, this explains the absence of the Cauchy horizon, while it exists for the $\beta = 0$ case. However, we did not try to control the Brans-Dicke field using a potential and hence further points to be clarified, this issue will be studied in a future paper \textit{II}, which will focus on mass inflation and dependence on parameters and potentials. However, the current paper suggests that there are are many other parameters and issues which will be interesting to study.

Although not a string-inspired model per se, the $\omega < -3/2$ and $\beta > 0$ cases are also interesting in the sense that a charged matter collapse can induce inflation, while the outside still does not inflate. So, as long as the theory allows for the existence of ghosts, the formation of a (at least one-way traversable) wormhole does not require complicated initial conditions. Of course, the existence of the ghost may be accompanied by complicated instabilities and this calls for further investigations. A detailed study of this issue is also postponed to a future paper.

To see, not only outside, but also inside the black hole and to study the detailed relations between gravity and various field contents, probably the best way is to use the double-null formalism. The authors hope that this paper will be an important step to prepare for upcoming investigations of diverse and various interesting topics, including the interests of theoretical gravity, string theory, holography, astrophysical applications, as well as cosmology, by varying matter, dimensions, gravity model, and symmetry, etc.

\newpage

\section*{Acknowledgments}
We would like to thank to Bum-Hoon Lee and also give thanks for using computer facilities in Center for Quantum Spacetime, Sogang University. DY would like to thank to Dong-il Hwang for help at the beginning of this project. DY was supported by the JSPS Grant-in-Aid for Scientific Research (A) No.~21244033 and Leung Center for Cosmology and Particle Astrophysics (LeCosPA) of National Taiwan University (103R4000). JH was supported in part by the programs of the Construction and Operation for Large-scale Science Data Center at KISTI (K-14-L01-C06-S01) and the Global Hub for Experiment Data of Basic Science, NRF (N-14-NM-IR06) and by the APCTP Topical Research Program.

\section*{Appendix A. Mass inflation in Einstein gravity}
In this appendix, we briefly review mass inflation in Einstein gravity. Mass inflation is a typical phenomenon of a deep inside of a charged black hole. Let us begin from a static charged black hole solution, so-called the Reissner-Nordstrom solution \cite{RN}:
\begin{eqnarray}
ds^{2} = -N^{2} dt^{2} + N^{-2}dr^{2} + r^2 d\Omega^2,
\end{eqnarray}
where $N^{2} = 1 - 2M/r + Q^{2}/r^{2}$. This solution has two horizons: $r_{\pm} = M \pm \sqrt{M^{2}-Q^{2}}$ with asymptotic mass $M$ and charge $Q$ (if $M > Q$). We show that the inner horizon is unstable up to a small perturbation, in the sense that a local observer near the inner horizon who sees a small perturbation can experience an exponential growth of the energy density.

For an in-going matter pulse, the energy-momentum tensor can be approximated by \cite{Poisson:1990eh,P}
\begin{eqnarray}
T_{\mu\nu} \propto \partial_{\mu}v \partial_{\nu}v \frac{F(v)}{r^{2}},
\end{eqnarray}
where $F(v)$ is so-called the luminosity function that should be proportional to $v^{-p}$ with a certain constant $p$ for a realistic matter \cite{Price}. Now, when there is an out-going observer who sees the in-going matter pulse, the observer measures the energy density $\rho = T_{\mu\nu} l^{\mu}l^{\nu}$, where $l^{\mu}$ is the null geodesic of the out-going observer. Note that $l^{\mu} \partial_{\mu} v = dv/d\eta \propto \exp \kappa_{-}v$, where $\kappa_{-}$ is the surface gravity of the inner horizon and $\eta$ is the affine parameter of the observer \cite{Poisson:1990eh,P}. Therefore, the energy density becomes
\begin{eqnarray}
\rho \propto e^{2 \kappa_{-} v}.
\end{eqnarray}
This is so-called mass inflation.

First, this implies that the inner horizon of the Reissner-Nordstrom solution is unstable via a small perturbation \cite{Poisson:1990eh}. Therefore, if we consider gravitational collapses, we cannot trust the Reissner-Nordstrom solution for the internal causal structure and the internal structure is dramatically changed \cite{BDIM}. This necessarily requires numerical simulations. Secondly, as the local energy function exponentially increases, other curvature quantities also exponentially increases \cite{Poisson:1990eh}. In the end, the inner horizon becomes a physical curvature singularity with a non-zero area, where the singularity is null for classical cases, while space-like with semi-classical back-reactions \cite{Yeom3}. In terms of double-null coordinates, mass inflation is usually identified by the behavior of the metric function $\alpha$ (when we use the double-null metric $ds^{2} = - \alpha^{2} dudv + r^{2} d\Omega^{2}$); $\alpha$ exponentially decreases, i.e., $\alpha^{2} \propto e^{-c v}$ with a constant $c$. This was repeatedly observed by numerical simulations \cite{doublenull,Yeom3}. In the extremely small $\alpha$ limit, all curvature quantities (Ricci scalar, Kretchmann scalar, etc.) behave like $\alpha^{-n}$ ($n>1$) and hence curvatures exponentially diverge.

All of these descriptions are on mass inflation in Einstein gravity. On the other hand, if there is a complicated coupling with charge, matter, and gravity, then these expectations can be modified and this is one important motivation of this paper.

\section*{Appendix B. Numerical convergence and consistency tests}
In this appendix, we demonstrate that 1) our numerical solutions are converging when including the effects of nontrivial parameters and scalar fields and 2) that the converged numerical solution is indeed a solution which satisfies the constraint equations.

The results presented here are very similar to those presented in \cite{Hansen:2013vha} which is not surprising, since the code in that paper and this one are essentially identical. For this reason we refer to that paper and references therein (especially, \cite{authors}), for further details of the numerical schemes and other details related to the design of the code.

The initial conditions and computational domain for the tests in this Appendix are similar to those used for making Figures~\ref{fig:charge_beta0} and \ref{fig:BD_beta0} for the $\omega=-1.4$, $e=0.3$, $\beta=0$ case. This setup is clearly non-trivial and is thus a good test for the general convergence properties of our code. Our computational domain for this convergence tests is, as for Figure~\ref{fig:causal}, in the range $v=[0;60]$ and $u=[0;40]$.

Next we demonstrate the convergence of the code by comparing a series of simulations with varying numerical resolution. For the tests in this Appendix, we do a total of six simulations, with each simulation changing the base resolution.

To limit the number of plots, we concentrate on displaying convergence results along the line $u=30$, which is well inside the apparent horizon and passes through a region with very complex behaviors (cf. Figures~\ref{fig:charge_beta0} and \ref{fig:BD_beta0}). Thus, if we see convergence here, it is a fair indication that convergence requirements are satisfied throughout the whole of the computational domain. 

Along this line ($u=30$), we calculate the relative convergence between two simulations (one with a numerical resolution twice that of the other) relative to a simulation with very high resolution:
\begin{equation}
  \label{eq:xidef}
  \xi \left( x_N^i \right) \equiv \frac{|x_N^i - x_{2N}^i|}{ |x_{\mathrm{HighRes}}^i |},
\end{equation}
where $x_N^i$ denotes the dynamic variable $x$ at the $i$-th grid point of simulation with resolution $N$ and where $x_{\mathrm{HighRes}}^i$ denotes the dynamic variable of the same $i$-th point for a simulation with the highest numerical resolution done by us. Obviously, this expression only makes sense for those $i$ points that coincide in all simulations.

\begin{figure}
\begin{center}
\includegraphics[scale=0.5]{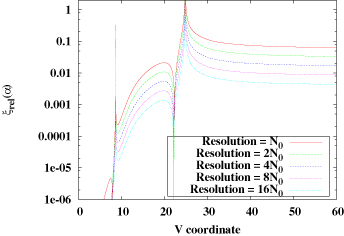}
\includegraphics[scale=0.5]{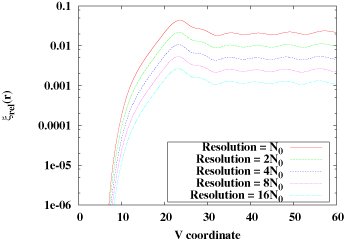}
\includegraphics[scale=0.5]{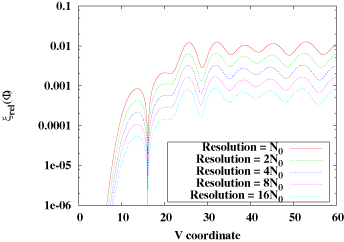}
\includegraphics[scale=0.5]{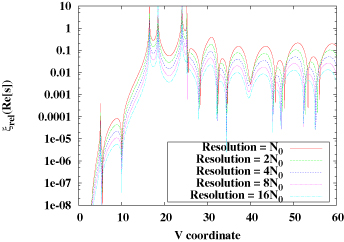}
\includegraphics[scale=0.5]{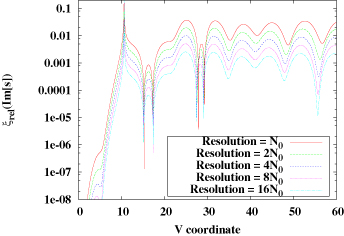}
\includegraphics[scale=0.5]{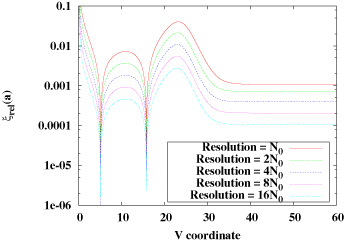}
\includegraphics[scale=0.5]{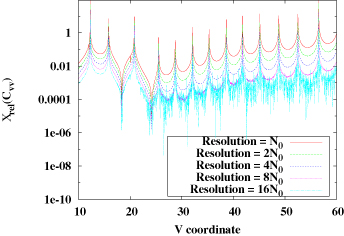}
\includegraphics[scale=0.5]{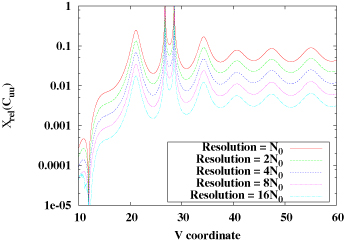}
\caption{\label{fig:convergence}Relative convergence of the dynamic variables and constraint
equations along line of $u=30$ for simulation with parameters: $\omega=-1.4$, $e=0.3$ and $\beta=0$. See text for details.}
\end{center}
\end{figure}

The first six plots in Figure~\ref{fig:convergence} show the relative convergence, $\xi$, for the dynamic variables $\alpha$, $r$, $\Phi$, Re~$s$, Im~$s$ and $a$, respectively. The lines in the figures are marked by their numerical resolution measured in terms of the most coarse resolution $N_0$, the high resolution simulation used to calculate expression Equation~(\ref{eq:xidef}), has a numerical resolution of $32$ times the base resolution, i.e., $N_{\mathrm{HighRes}}=32N_0$.

From these figures, it is clear that the dynamic variables are converging for simulations of increasing
resolution. Furthermore, since we plot the \textit{relative} convergence of the dynamic variables, we see that the relative
change between the two highest resolution simulations show that the variables change on the order of $1~\%$ or less, which we considered a quite acceptable. 

It is, however, not enough to demonstrate that the simulations are converging, they must also converge to a `\textit{physical}' solution, i.e., the residuals of the constraint equations (Equations~(\ref{eq:E1}) and (\ref{eq:E2})) must converge to zero. To demonstrate this, we calculate the relative convergence of the constraint equation residuals, (relative to the Einstein-tensor) in a similar way to Equation~(\ref{eq:xidef}):
\begin{equation}
  \label{eq:chidef}
  \chi \left( C_N^i \right) \equiv \frac{|C_N^i|}{ |G_{\mathrm{HighRes}}^i |},
\end{equation}
where $C_N^i$ denotes the residual of the constraint equation ($C_{uu}$ or $C_{vv}$), at the $i$-th point for simulation with
resolution $N$ and where $G_{\mathrm{HighRes}}^i$ denotes the corresponding Einstein-tensor component ($G_{uu}$ or $G_{vv}$ respectively) at the same point.

The relative convergence of the residuals of the constraint equations are demonstrated in the bottom two plots of Figure~\ref{fig:convergence}, where it is seen that they converge towards zero for higher resolution simulations. This indicates that not only are the numerical solutions converging for simulations of higher resolution, but that they are indeed converging towards a physical solution. 

Finally, it is emphasized that the results presented in this appendix are not the only convergence tests that we have
performed, they are merely representative of typical convergence behavior. For all results presented in
this paper, we have performed a large number of simulations with varying resolutions to ensure that the results had converged to their physical solution.

\section*{Appendix C. Catalog of metric function and energy-momentum tensors}

\begin{figure}
\begin{center}
\includegraphics[scale=0.17]{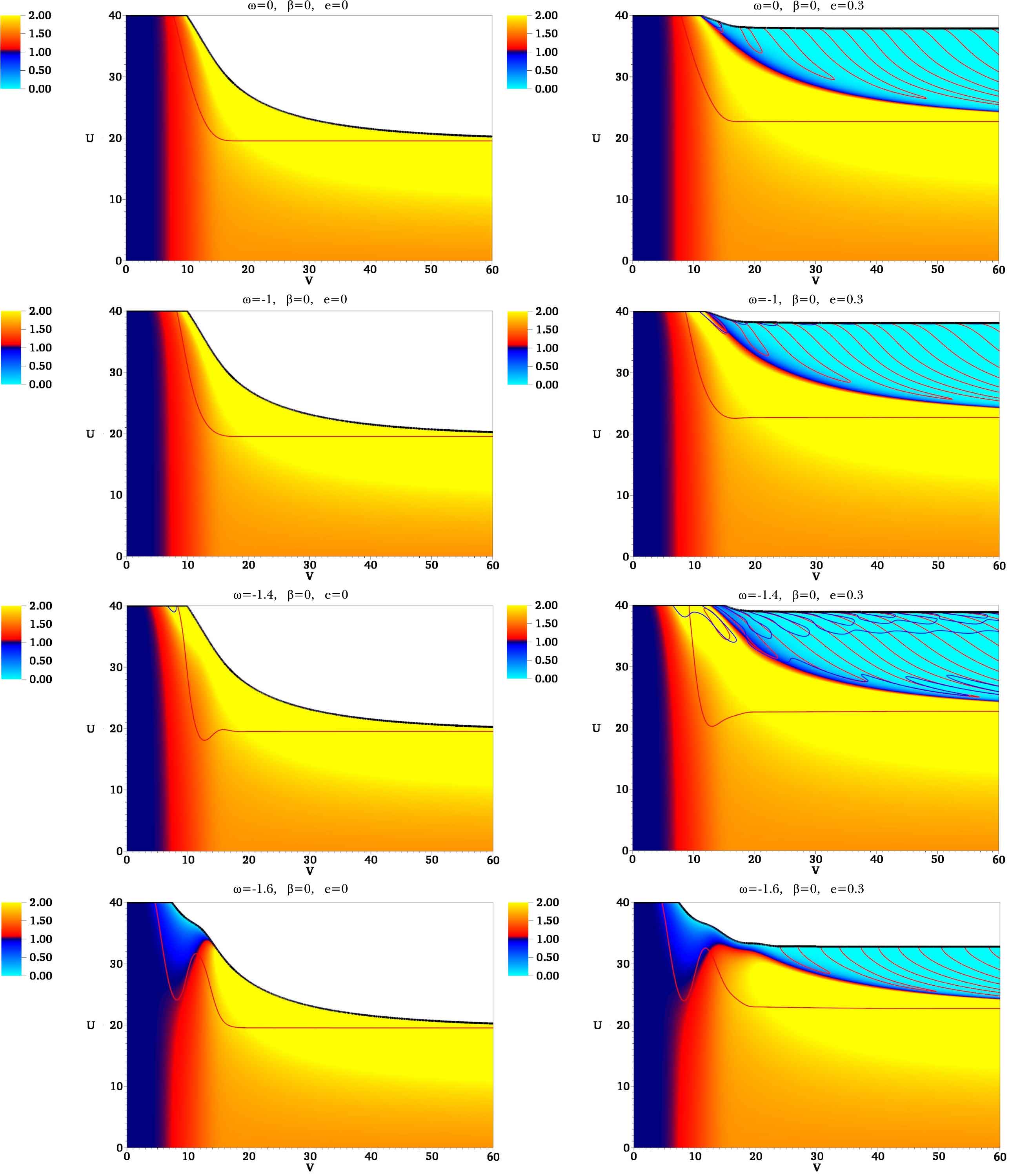}
\caption{\label{fig:alpha_beta0}The plots of $\alpha$ for $\beta=0$.}
\end{center}
\end{figure}

\begin{figure}
\begin{center}
\includegraphics[scale=0.17]{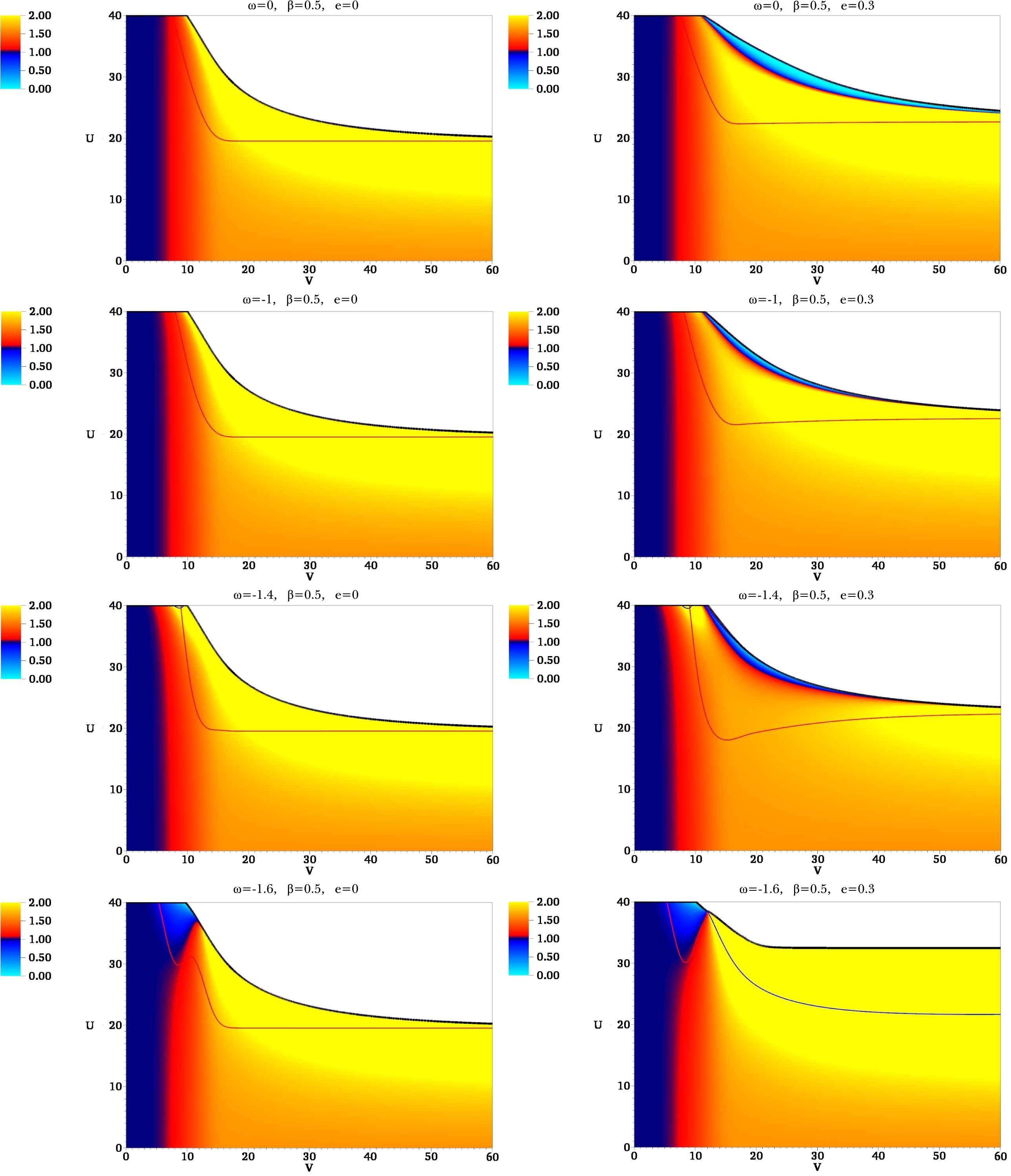}
\caption{\label{fig:alpha_beta05}The plots of $\alpha$ for $\beta=0.5$.}
\end{center}
\end{figure}

\begin{figure}
\begin{center}
\includegraphics[scale=0.17]{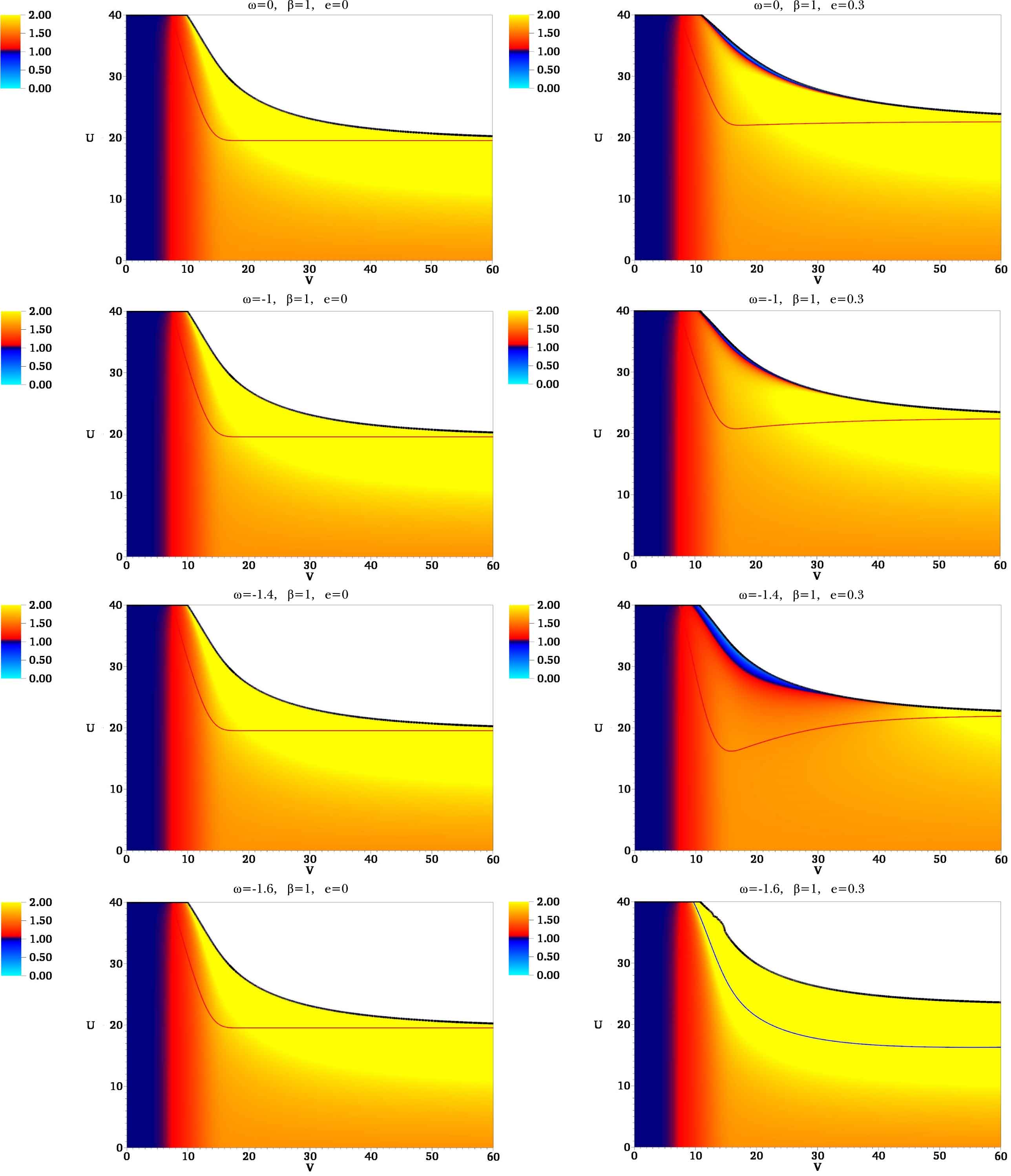}
\caption{\label{fig:alpha_beta1}The plots of $\alpha$ for $\beta=1$.}
\end{center}
\end{figure}

In this appendix, we summarize the metric function $\alpha$ and energy-momentum tensor components $T_{uu}$ and $T_{vv}$. Figures~\ref{fig:alpha_beta0}, \ref{fig:alpha_beta05}, and \ref{fig:alpha_beta1} are plots of $\alpha$ for $\beta = 0$, $0.5$, and $1$, respectively. For typical situations of mass inflation, $\alpha \rightarrow 0$ exponentially, and hence almost all curvature quantities (Ricci scalar, Kretchmann scalar, etc.) behave $\alpha^{-n}$ ($n>1$) and hence curvatures exponentially diverge.

\begin{figure}
\begin{center}
\includegraphics[scale=0.17]{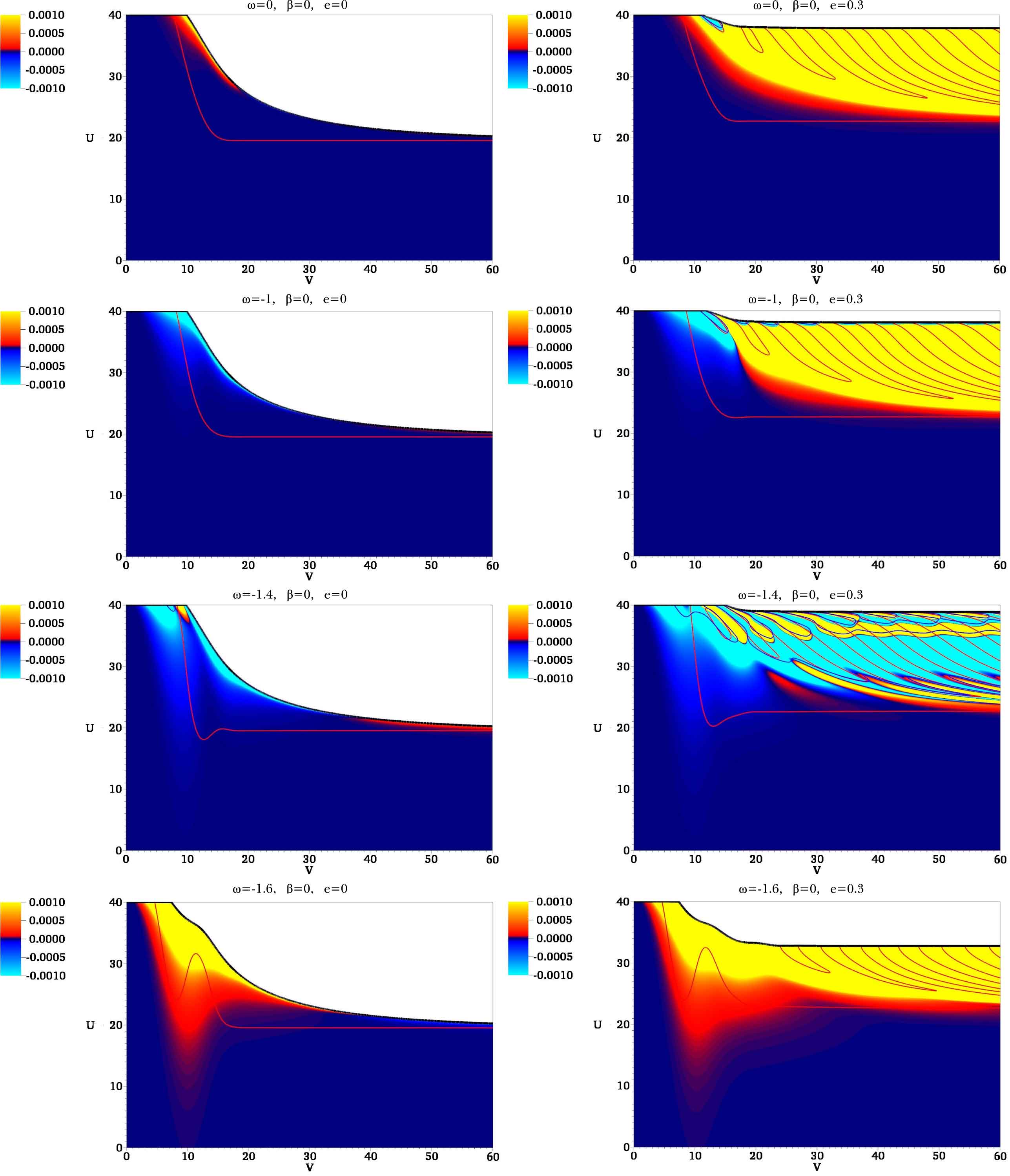}
\caption{\label{fig:Tuu_beta0}The plots of $T_{uu}$ for $\beta=0$.}
\end{center}
\end{figure}

\begin{figure}
\begin{center}
\includegraphics[scale=0.17]{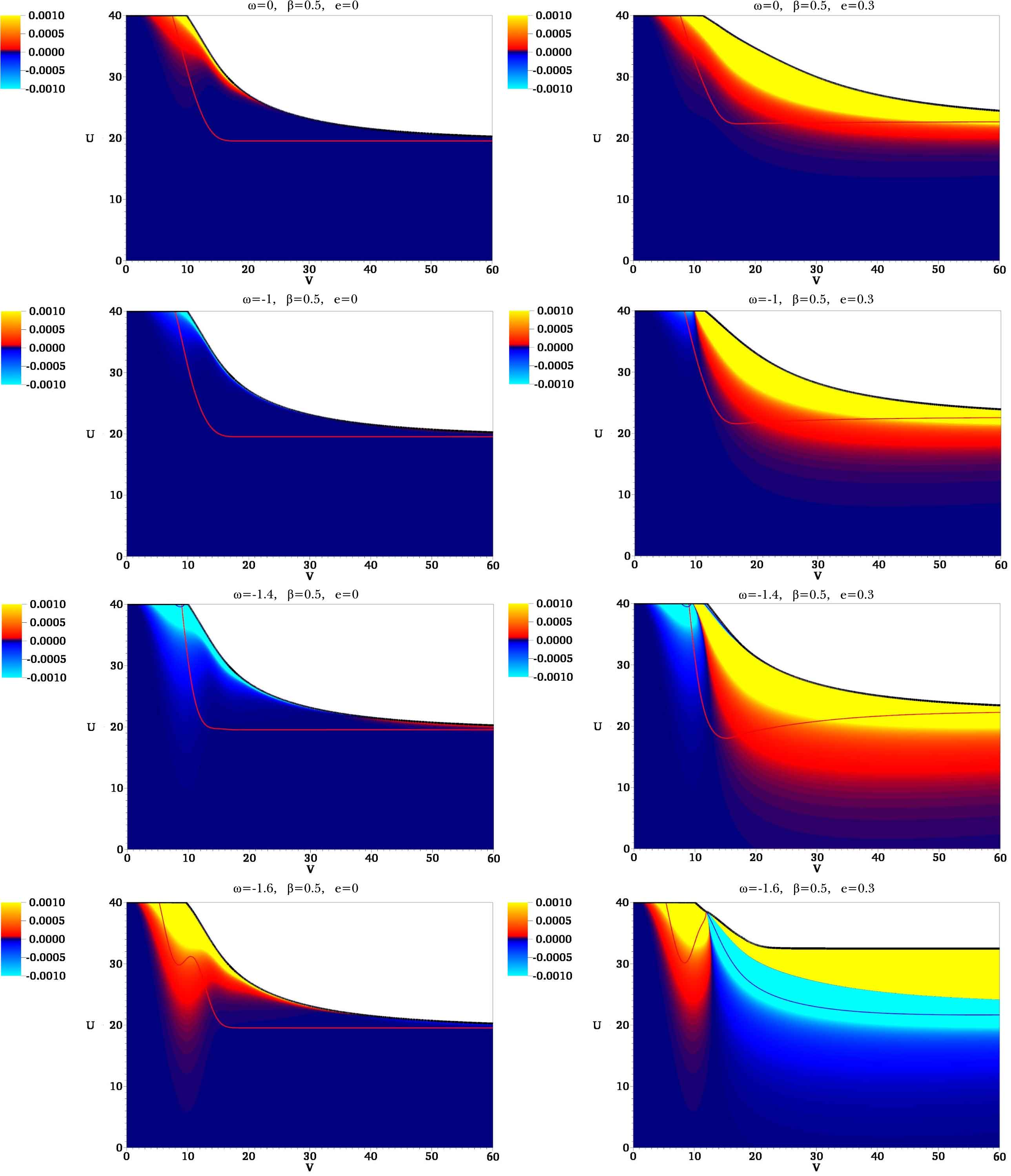}
\caption{\label{fig:Tuu_beta05}The plots of $T_{uu}$ for $\beta=0.5$.}
\end{center}
\end{figure}

\begin{figure}
\begin{center}
\includegraphics[scale=0.17]{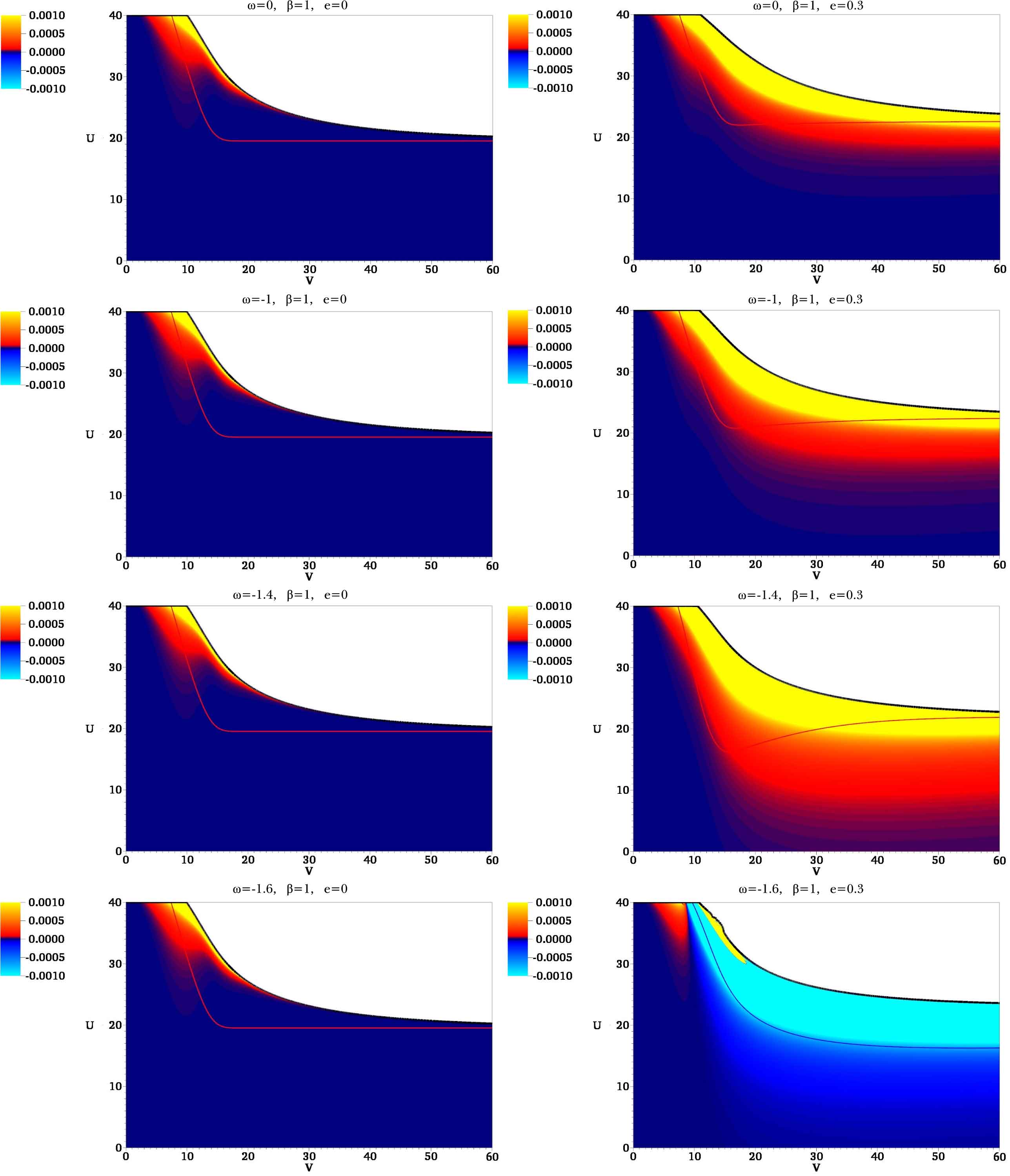}
\caption{\label{fig:Tuu_beta1}The plots of $T_{uu}$ for $\beta=1$.}
\end{center}
\end{figure}

\begin{figure}
\begin{center}
\includegraphics[scale=0.17]{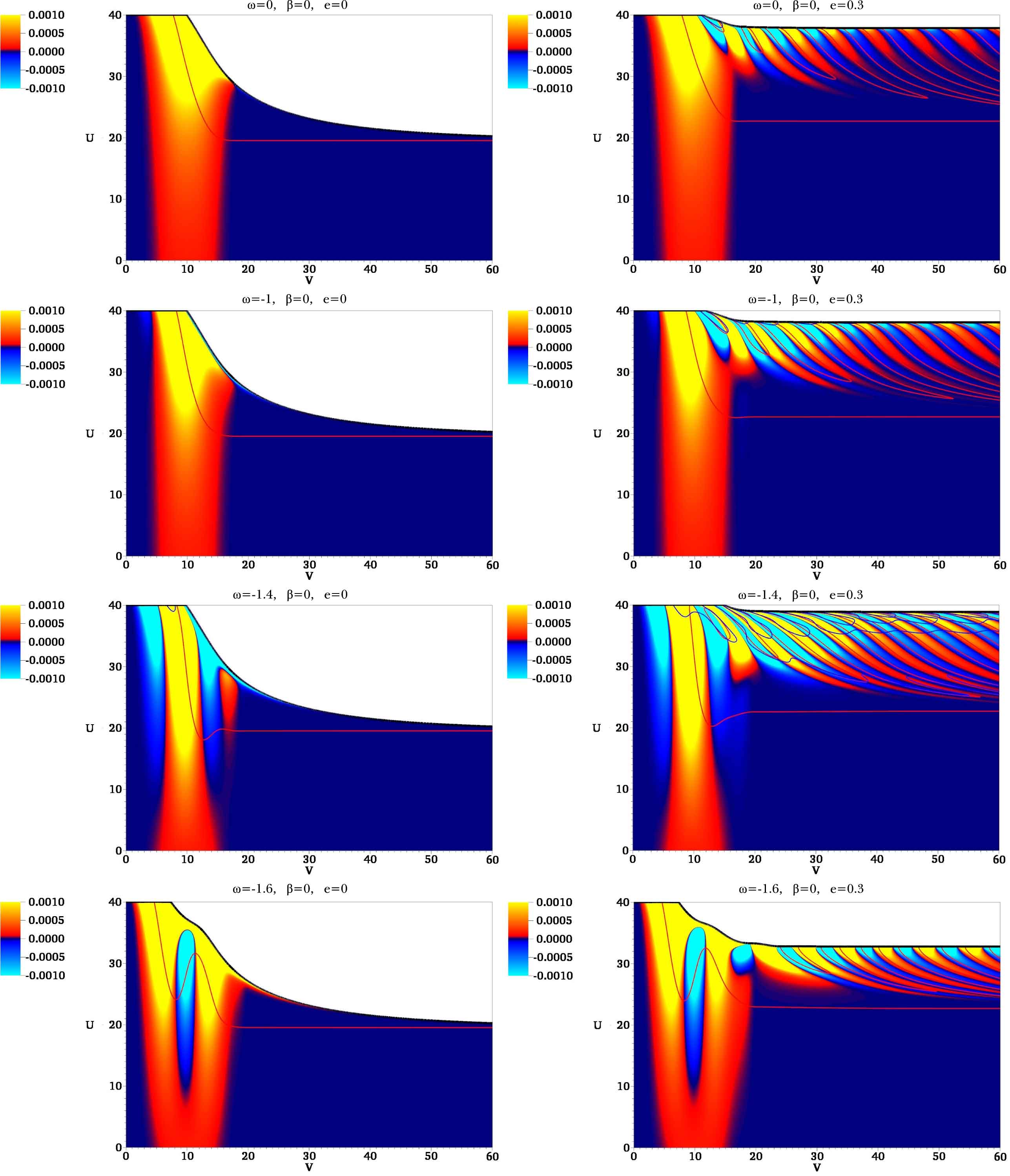}
\caption{\label{fig:Tvv_beta0}The plots of $T_{vv}$ for $\beta=0$.}
\end{center}
\end{figure}

\begin{figure}
\begin{center}
\includegraphics[scale=0.17]{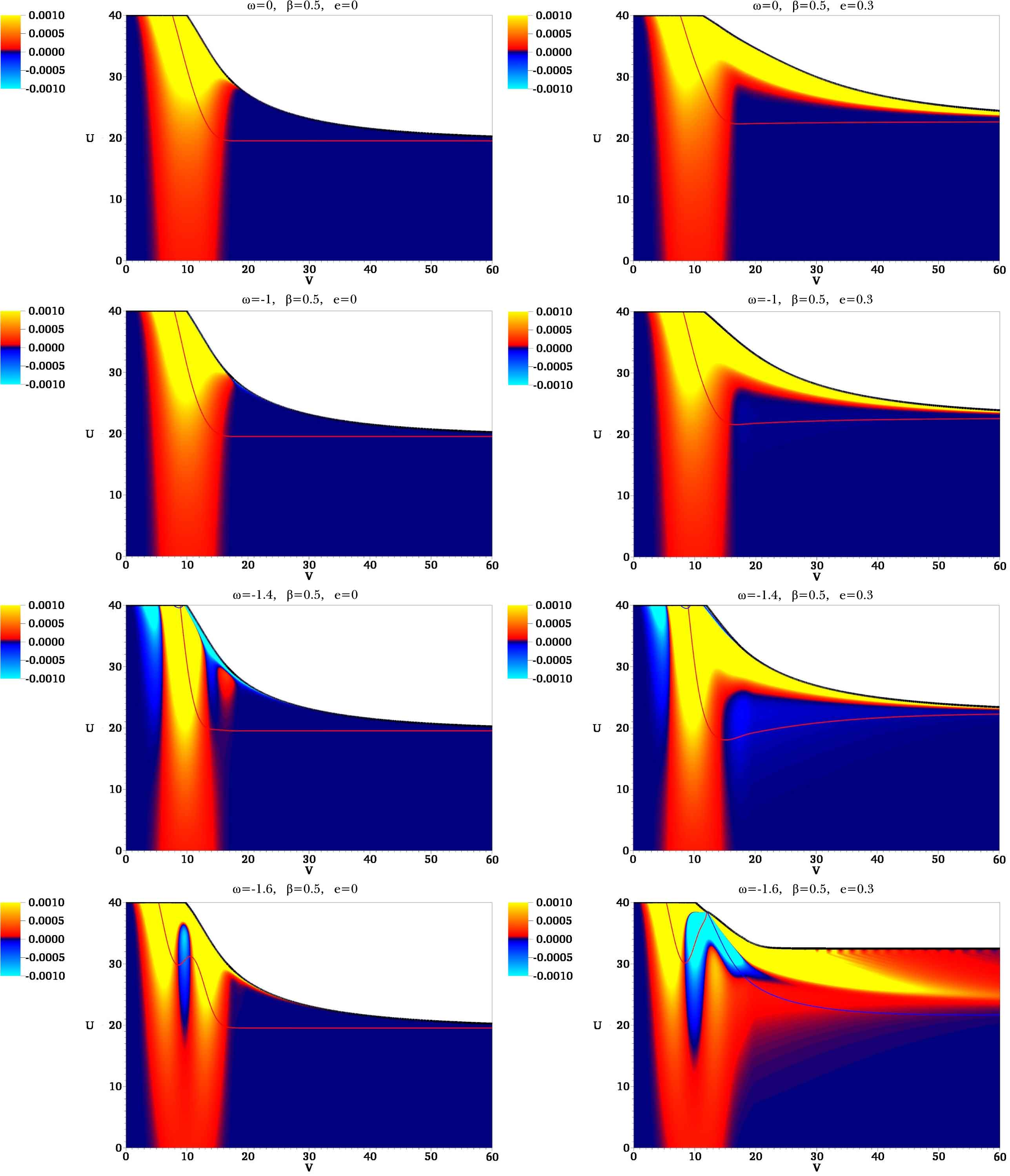}
\caption{\label{fig:Tvv_beta05}The plots of $T_{vv}$ for $\beta=0.5$.}
\end{center}
\end{figure}

\begin{figure}
\begin{center}
\includegraphics[scale=0.17]{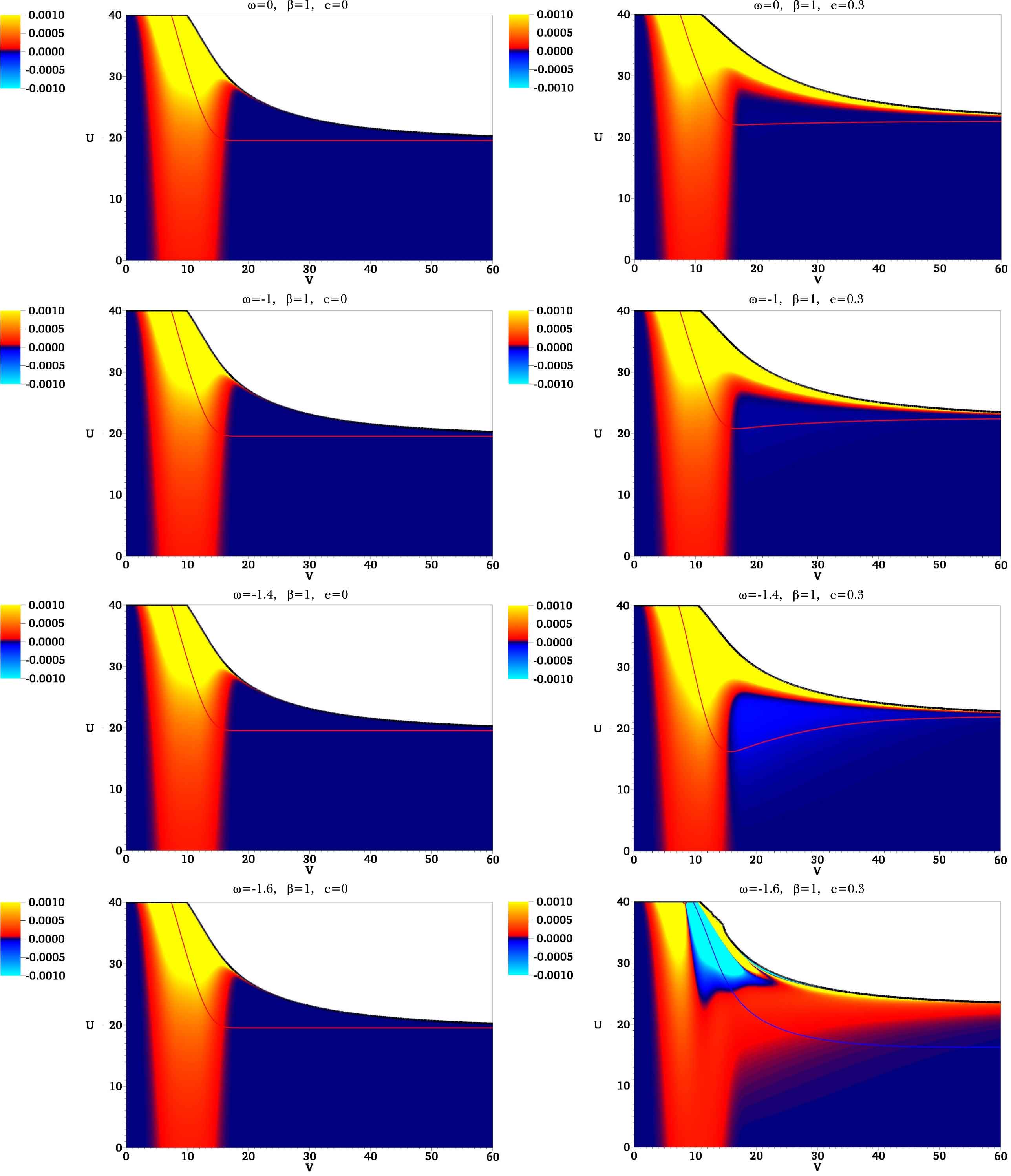}
\caption{\label{fig:Tvv_beta1}The plots of $T_{vv}$ for $\beta=1$.}
\end{center}
\end{figure}

Figures~\ref{fig:Tuu_beta0}, \ref{fig:Tuu_beta05}, and \ref{fig:Tuu_beta1} are plots of $T_{uu}$ for $\beta = 0$, $0.5$, and $1$, respectively. Figures~\ref{fig:Tvv_beta0}, \ref{fig:Tvv_beta05}, and \ref{fig:Tvv_beta1} are plots of $T_{vv}$ for $\beta = 0$, $0.5$, and $1$, respectively. Note that around the apparent horizon $r_{,v} = 0$,
\begin{eqnarray}
\left. r_{,vv} \right|_{r_{,v}=0} = - 4 \pi r T_{vv},
\end{eqnarray}
and hence, for an out-going observer, if $T_{vv} > 0$, then the $r_{,v}=0$ horizon changes the sign of $r_{,v}$ from $+$ to $-$, while $T_{vv} < 0$, changes the sign of $r_{,v}$ from $-$ to $+$. Also, around the apparent horizon $r_{,u}=0$,
\begin{eqnarray}
\left. r_{,uu} \right|_{r_{,u}=0} = - 4 \pi r T_{uu},
\end{eqnarray}
and hence, for an in-going observer, if $T_{uu} > 0$, then the $r_{,u}=0$ horizon changes the sign of $r_{,u}$ from $+$ to $-$, while $T_{uu} < 0$, changes the sign of $r_{,u}$ from $-$ to $+$. For all figures, these rules are always satisfied even with complicated horizon dynamics. This is also a simple check of the consistency of our simulations.

\newpage

\begin{thebibliography}{200}

\bibitem{DeWitt:1967yk}
  B.~S.~DeWitt,
  Phys.\ Rev.\  {\bf 160}, 1113 (1967);\\
  B.~S.~DeWitt,
  Phys.\ Rev.\  {\bf 162}, 1195 (1967); \\
  B.~S.~DeWitt,
  Phys.\ Rev.\  {\bf 162}, 1239 (1967).

\bibitem{Kiefer}
  C. Kiefer, ``{\it Quantum gravity,}'' Oxford University Press (2004);\\
  C.~Rovelli, ``{\it Quantum gravity,}'' Cambridge University Press (2004);\\
  G.~W.~Gibbons and S.~W.~Hawking, ``{\it Euclidean quantum gravity,}'' World Scientific (1993).

\bibitem{Green:1987sp}
  M.~B.~Green, J.~H.~Schwarz and E.~Witten,
  ``{\it Superstring theory. Vol. 1: Introduction,}'' Cambridge University Press (1987); \\
  M.~B.~Green, J.~H.~Schwarz and E.~Witten,
  ``{\it Superstring theory. Vol. 2: Loop amplitudes, anomalies and phenomenology,}'' Cambridge University Press (1987).

\bibitem{Becker:2007zj} 
  K.~Becker, M.~Becker and J.~H.~Schwarz,
  ``{\it String theory and M-theory: A modern introduction,}'' Cambridge University Press (2007).

\bibitem{Kachru:2003aw}
  S.~Kachru, R.~Kallosh, A.~D.~Linde and S.~P.~Trivedi,
  Phys.\ Rev.\  D {\bf 68}, 046005 (2003)
  [arXiv:hep-th/0301240];\\
  J.~P.~Conlon,
  Fortsch.\ Phys.\  {\bf 55}, 287 (2007)
  [hep-th/0611039].

\bibitem{Maldacena:1997re}
  J.~M.~Maldacena,
  Adv.\ Theor.\ Math.\ Phys.\  {\bf 2}, 231 (1998)
  [Int.\ J.\ Theor.\ Phys.\  {\bf 38}, 1113 (1999)]
  [arXiv:hep-th/9711200].

\bibitem{Conlon:2010jq} 
  J.~P.~Conlon and F.~G.~Pedro,
  JHEP {\bf 1105}, 079 (2011)
  [arXiv:1010.2665 [hep-th]];\\
  D.~Hwang, F.~G.~Pedro and D.~Yeom,
  JHEP {\bf 1309}, 159 (2013)
  [arXiv:1306.6687 [hep-th]].

\bibitem{Hawking:1969sw}
  S.~W.~Hawking and R.~Penrose,
  Proc.\ Roy.\ Soc.\ Lond.\  A {\bf 314}, 529 (1970).

\bibitem{Borde:2001nh}
  A.~Borde, A.~H.~Guth and A.~Vilenkin,
  Phys.\ Rev.\ Lett.\  {\bf 90}, 151301 (2003)
  [arXiv:gr-qc/0110012].

\bibitem{Gasperini:2007zz}
  M.~Gasperini, ``{\it Elements of string cosmology,}'' Cambridge University Press (2007).

\bibitem{Brans:1961sx}
  C.~Brans and R.~H.~Dicke,
  Phys.\ Rev.\  {\bf 124}, 925 (1961).

\bibitem{Fujii:2003pa}
  Y.~Fujii and K.~Maeda, ``{\it The scalar-tensor theory of gravitation,}'' Cambridge University Press (2003);\\
  V.~Faraoni, ``{\it Cosmology in scalar tensor gravity,}'' Kluwer Academic Publishers (2004).

\bibitem{Gibbons:1987ps} 
  G.~W.~Gibbons and K.~-i.~Maeda,
  Nucl.\ Phys.\ B {\bf 298}, 741 (1988);\\
  D.~Garfinkle, G.~T.~Horowitz and A.~Strominger,
  Phys.\ Rev.\ D {\bf 43}, 3140 (1991)
  [Erratum-ibid.\ D {\bf 45}, 3888 (1992)].

\bibitem{Ashtekar:2004cn}
  A.~Ashtekar and B.~Krishnan,
  Living Rev.\ Rel.\ {\bf 7}, 10 (2004) [arXiv:gr-qc/0407042]; \\
  S.~A.~Hayward,
  arXiv:gr-qc/9303006; \\
  A.~B.~Nielsen and D.~Yeom,
  Int.\ J.\ Mod.\ Phys.\  A {\bf 24}, 5261 (2009) [arXiv:0804.4435 [gr-qc]].

\bibitem{Poisson:1990eh}
  E.~Poisson and W.~Israel,
  Phys.\ Rev.\  D {\bf 41}, 1796 (1990);\\
  A.~Ori,
  Phys.\ Rev.\ Lett.\  {\bf 67}, 789 (1991); \\
  A.~Ori,
  Phys.\ Rev.\ Lett.\  {\bf 68}, 2117 (1992).

\bibitem{Yeom2}
  D.~Hwang and D.~Yeom,
  Class.\ Quant.\ Grav.\  {\bf 27}, 205002 (2010)
  [arXiv:1002.4246 [gr-qc]].

\bibitem{Israel:1967wq} 
  W.~Israel,
  Phys.\ Rev.\  {\bf 164}, 1776 (1967);\\
  W.~Israel,
  Commun.\ Math.\ Phys.\  {\bf 8}, 245 (1968).

\bibitem{Hamade:1995ce}
  R.~S.~Hamade and J.~M.~Stewart,
  Class.\ Quant.\ Grav.\  {\bf 13}, 497 (1996)
  [arXiv:gr-qc/9506044].

\bibitem{doublenull}
  R.~Parentani and T.~Piran, Phys.\ Rev.\ Lett {\bf 73}, 2805 (1994) [arXiv:hep-th/9405007];\\
  T.~Chiba and J.~Soda, Prog.\ Theor.\ Phys.\  {\bf 96}, 567 (1996) [arXiv:gr-qc/9603056]; \\
  S.~Ayal and T.~Piran, Phys.\ Rev.\  D {\bf 56}, 4768 (1997) [arXiv:gr-qc/9704027];\\
  S.~Hod and T.~Piran, Phys.\ Rev.\ Lett {\bf 81}, 1554 (1998) [arXiv:gr-qc/9803004];\\
  S.~Hod and T.~Piran, Gen.\ Rel.\ Grav.\  {\bf 30}, 1555 (1998) [arXiv:gr-qc/9902008];\\
  E.~Sorkin and T.~Piran, Phys.\ Rev.\ D {\bf 63}, 084006 (2001) [arXiv:gr-qc/0009095];\\
  E.~Sorkin and T.~Piran, Phys.\ Rev.\  D {\bf 63}, 124024 (2001) [arXiv:gr-qc/0103090];\\
  Y.~Oren and T.~Piran, Phys.\ Rev.\ D {\bf 68}, 044013 (2003) [arXiv:gr-qc/0306078];\\
  P.~P.~Avelino, A.~J.~S.~Hamilton and C.~A.~R.~Herdeiro, Phys.\ Rev.\  D {\bf 79}, 124045 (2009) [arXiv:0904.2669 [gr-qc]].

\bibitem{Ann}   
  A.~Borkowska, M.~Rogatko and R.~Moderski, Phys.\ Rev.\  D {\bf 83}, 084007 (2011) [arXiv:1103.4808 [hep-th]];\\
  A.~Nakonieczna, M.~Rogatko and R.~Moderski,  Phys.\ Rev.\ D {\bf 86}, 044043 (2012)  [arXiv:1209.1203 [hep-th]];\\
  A.~Nakonieczna and M.~Rogatko,  arXiv:1209.3614 [hep-th].
  
\bibitem{Scheel:1994yr}
  M.~A.~Scheel, S.~L.~Shapiro and S.~A.~Teukolsky,
  Phys.\ Rev.\  D {\bf 51}, 4208 (1995) [arXiv:gr-qc/9411025]; \\
  M.~A.~Scheel, S.~L.~Shapiro and S.~A.~Teukolsky,
  Phys.\ Rev.\  D {\bf 51}, 4236 (1995) [arXiv:gr-qc/9411026].

\bibitem{authors}
  J.~Hansen, A.~Khokhlov and I.~Novikov, Phys.\ Rev.\  D {\bf 71}, 064013 (2005) [arXiv:gr-qc/0501015];\\
  A.~Doroshkevich, J.~Hansen, I.~Novikov and A.~Shatskiy, Int. J. Mod. Phys. D {\bf 18}, 11 (2009) [arXiv:0812.0702 [gr-qc]];\\
  A.~Doroshkevich, J.~Hansen, D.~Novikov, I.~Novikov, D. H.~Park and A.~Shatskiy, Phys.\ Rev.\  D {\bf 81}, 124011 (2010)
  [arXiv:0908.1300 [gr-qc]].

\bibitem{Yeom1}
  D.~Yeom,
  arXiv:0912.0068 [gr-qc];\\
  D.~Hwang and D.~Yeom,
  Class.\ Quant.\ Grav.\  {\bf 28}, 155003 (2011)
  [arXiv:1010.3834 [gr-qc]];\\
  B.~-H.~Lee and D.~Yeom,
  Nuovo Cim.\ C {\bf 036}, no. s01, 79 (2013)
  [arXiv:1111.0139 [gr-qc]];\\
  D.~Hwang, B.~-H.~Lee and D.~Yeom,
  JCAP {\bf 1301}, 005 (2013)
  [arXiv:1210.6733 [gr-qc]].

\bibitem{Yeom3}
  S.~E.~Hong, D.~Hwang, E.~D.~Stewart and D.~Yeom,
  Class.\ Quant.\ Grav.\  {\bf 27}, 045014 (2010) [arXiv:0808.1709 [gr-qc]];\\
  D.~Hwang and D.~Yeom,
  Phys.\ Rev.\ D {\bf 84}, 064020 (2011)
  [arXiv:1010.2585 [gr-qc]].

\bibitem{Yeom5}
  D.~Hwang, B.~-H.~Lee and D.~Yeom,
  JCAP {\bf 1112}, 006 (2011)
  [arXiv:1110.0928 [gr-qc]].

\bibitem{Hansen:2013vha} 
  J.~Hansen, B.~-H.~Lee, C.~Park and D.~Yeom,
  Class.\ Quant.\ Grav.\  {\bf 30}, 235022 (2013)
  [arXiv:1307.0266 [hep-th]].

\bibitem{Hwang:2010aj}
  D.~Hwang, H.~-B.~Kim and D.~Yeom,
  Class.\ Quant.\ Grav.\  {\bf 29}, 055003 (2012)
  [arXiv:1105.1371 [gr-qc]].

\bibitem{Yeom4}
  D.~Hwang, B.~-H.~Lee, W.~Lee and D.~Yeom,
  JCAP {\bf 1207}, 003 (2012)
  [arXiv:1201.6109 [gr-qc]].

\bibitem{Foffa:1999dv}
  S.~Foffa, M.~Maggiore and R.~Sturani,
  Nucl.\ Phys.\  B {\bf 552}, 395 (1999)
  [arXiv:hep-th/9903008].

\bibitem{Randall:1999ee}
  L.~Randall and R.~Sundrum,
  Phys.\ Rev.\ Lett.\  {\bf 83}, 3370 (1999) [arXiv:hep-ph/9905221].

\bibitem{Garriga:1999yh}
  J.~Garriga and T.~Tanaka,
  Phys.\ Rev.\ Lett.\  {\bf 84}, 2778 (2000) [arXiv:hep-th/9911055].

\bibitem{Sotiriou:2008rp}
  T.~P.~Sotiriou and V.~Faraoni,
  Rev.\ Mod.\ Phys.\  {\bf 82}, 451 (2010)
  [arXiv:0805.1726 [gr-qc]];\\
  S.~Nojiri and S.~D.~Odintsov,
  Int.\ J.\ Geom.\ Meth.\ Mod.\ Phys.\  {\bf 4}, 115 (2007)
  [arXiv:hep-th/0601213].

\bibitem{Agnese:1995kd}
  A.~G.~Agnese and M.~La Camera,
  Phys.\ Rev.\  D {\bf 51}, 2011 (1995).

\bibitem{JHY}
  J.~Hansen, D.~Hwang and D.~Yeom,
  JHEP {\bf 0911}, 016 (2009)
  [arXiv:0908.0283 [gr-qc]].

\bibitem{Sotiriou:2011dz} 
  T.~P.~Sotiriou and V.~Faraoni,
  Phys.\ Rev.\ Lett.\  {\bf 108}, 081103 (2012)
  [arXiv:1109.6324 [gr-qc]].

\bibitem{RN}
H. Reissner, Ann. Physik., {\bf 50}, 106 (1916);\\
G. Nordstrom, Proc. Kon. Ned. Akad. Wet., {\bf 20}, 1238 (1918).

\bibitem{P}
E. Poisson, [arXiv:gr-qc/9709022].

\bibitem{Price}
R.H. Price, Phys. Rev. D {\bf 5}, 2419 (1972);\\
R.H. Price, Phys. Rev. D {\bf 5}, 2439 (1972).

\bibitem{BDIM}
A. Bonanno, S. Droz, W. Israel and S. M. Morsink, Phys. Rev. D {\bf 50}, 7372 (1994)
[arXiv:gr-qc/9403019].

\end{thebibliography}
\end{document}